\documentclass{article}
\pdfoutput=1

\usepackage{PRIMEarxiv}

\usepackage[utf8]{inputenc} %
\usepackage[T1]{fontenc}    %
\usepackage[expansion=false]{microtype}
\usepackage{amsmath,amssymb,amsfonts,mathtools}
\usepackage{bm}
\usepackage{url}            %
\usepackage{booktabs}       %
\usepackage{amsfonts}       %
\usepackage{nicefrac}       %
\usepackage{microtype}      %
\usepackage{fancyhdr}       %
\usepackage[table]{xcolor}
\usepackage{graphicx}       %
\graphicspath{{media/}}     %

\usepackage{multirow}
\usepackage{makecell}
\usepackage{placeins}
\usepackage{array}
\usepackage{tabularx}
\usepackage{enumitem}
\usepackage[numbers,sort&compress]{natbib}
\usepackage[colorlinks=true,linkcolor=blue,citecolor=blue,urlcolor=blue]{hyperref}
\usepackage{caption}
\usepackage{tikz}
\usetikzlibrary{arrows.meta,positioning,calc,decorations.pathreplacing}

\newcommand{\luna}{Luna-TTS Family}
\newcommand{\lunaNAR}{Luna-TTS}
\newcommand{\lunaBlock}{Luna-TTS Realtime}
\newcommand{\lunaTok}{Luna-Codec}          %

\newcommand{\dataScale}{1 million hours}          %
\newcommand{\dataScaleShort}{1M hours}
\newcommand{\tokDataScale}{400K hours}              %
\newcommand{\langList}{Chinese, English, Japanese, and Korean}
\newcommand{\numLangs}{four}

\newcommand{\modelSize}{0.6B}                        %
\newcommand{\initLLM}{Qwen3-0.6B}                    %
\newcommand{\initLLMFamily}{Qwen3}                   %
\newcommand{\durLLM}{Qwen3-0.6B}                     %

\newcommand{\sampleRate}{24\,kHz}                    %
\newcommand{\frameRate}{25\,Hz}                      %
\newcommand{\numCodebooks}{8}                        %
\newcommand{\codebookSize}{2048}                     %
\newcommand{\bitrate}{2.2\,kbps}                     %
\newcommand{\tokPerSec}{200}                         %

\newcommand{\seedZhCER}{0.73}      %
\newcommand{\seedZhSIM}{79.7}      %
\newcommand{\seedEnWER}{1.49}      %
\newcommand{\seedEnSIM}{76.8}      %
\newcommand{\cvZhWER}{3.17}        %
\newcommand{\cvEnWER}{3.18}        %
\newcommand{\cvJaWER}{5.00}        %
\newcommand{\cvKoWER}{5.93}        %
\newcommand{\cvHardZhWER}{6.90}    %
\newcommand{\cvHardEnWER}{6.18}    %
\newcommand{\cvAvgWER}{4.32}       %
\newcommand{\cvAvgSIM}{73.5}       %

\newcommand{\demoURL}{\url{https://vuilabs-ai.github.io/luna-tts}}

\newcommand{\narSteps}{32}                           %
\newcommand{\blockFrames}{32}                        %
\newcommand{\blockMs}{1.28s}                         %
\newcommand{\blockSteps}{8}                          %

\definecolor{LunaHighlight}{HTML}{E8F6F0}

\title{\luna{} Technical Report
}

\author{
  VUI Labs Research
  \\[0.8em]
  \normalsize
  \raisebox{-0.18\height}{%
    \includegraphics[height=0.90em]{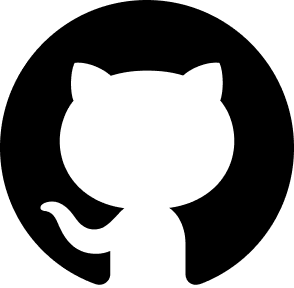}
  }%
  \quad \demoURL{}
}

\begin{document}
\maketitle
\begin{abstract}
Modern text-to-speech (TTS) is dominated by autoregressive (AR) codec language models, whose left-to-right decoding carries structural costs: latency that grows with utterance length, error accumulation along the committed prefix, and an artificial generation order imposed on the Residual Vector Quantization (RVQ) token grid, which possesses none. We propose \luna{}, a family of diffusion-language-model-based TTS systems pretrained on \dataScale{} of speech across \langList{}. The family is built by progressive adaptation of a pretrained AR text LLM, from causal to bidirectional and finally to block-causal attention, and comprises two variants sharing a single tokenizer, data pipeline, and \modelSize{} backbone lineage. \lunaNAR{} is fully non-autoregressive: it generates the entire RVQ token grid of an utterance in a fixed number of parallel refinement steps, with zero-shot voice cloning and speech editing arising natively as infilling. \lunaBlock{}, derived from \lunaNAR{} by continual training, is autoregressive over blocks of \blockFrames{} codec frames (\blockMs{}) while denoising each block in parallel; it supports KV-cached blockwise generation and incremental audio delivery after the first block is committed. Under the warmed local serving protocol, \lunaBlock{} achieves an end-to-end RTF of 0.0240 and commits its first \blockMs{} decoded audio block in 41.6\,ms, corresponding to more than $40\times$ real-time generation within the measured engine boundary. An annealed fine-tuning stage adds explicit control over emotion and non-verbal vocalizations, and a reinforcement-learning stage applies GRPO, with policy ratios computed over the realized denoising trajectory. On Seed-TTS-Eval, \lunaNAR{} achieves the best results on all four metrics among the compared open-source and commercial systems (\seedZhCER{} CER / \seedZhSIM{} SIM on test-zh, \seedEnWER{} WER / \seedEnSIM{} SIM on test-en); on the harder in-the-wild CV3-Eval, it likewise posts the lowest Mandarin and English error rates in our comparison. In expressive-control evaluations against leading commercial systems, it further achieves the best results on most objective, model-based, and human-rated metrics for non-verbal vocalization and emotion control.
\end{abstract}

\section{Introduction}
\label{sec:introduction}

Text-to-speech (TTS) synthesis has been reshaped by the language-modeling paradigm. By discretizing speech with neural audio codecs~\cite{zeghidour2021soundstream,defossez2022encodec,kumar2023dac} and training decoder-only Transformers with a next-token-prediction objective, neural codec language models~\cite{betker2023tortoise,wang2023valle} turned speech synthesis into a sequence-modeling problem that inherits the scaling behavior, in-context learning ability, and infrastructure of large language models (LLMs). Scaling this recipe along data, parameters, and post-training has produced a generation of systems that achieve human-parity naturalness and robust zero-shot voice cloning from a few seconds of reference audio, including Seed-TTS~\cite{anastassiou2024seedtts}, the CosyVoice series~\cite{du2024cosyvoice,du2024cosyvoice2,du2025cosyvoice3}, MiniMax-Speech~\cite{minimax2025speech}, Llasa~\cite{ye2025llasa}, GLM-TTS~\cite{cui2025glmtts}, Qwen3-TTS and Qwen-Audio-3.0-TTS~\cite{hu2026qwen3tts,xiang2026qwenaudio3tts}, Fish Audio S2~\cite{liao2026fishs2}, MOSS-TTS~\cite{gong2026mosstts}, and others~\cite{wang2025sparktts,zhou2025indextts2,guo2024fireredtts,xiaomi2025mimoaudio,stepfun2025stepaudio2}. In general, all of these systems share a common computational core: an \emph{autoregressive} (AR) language model that emits speech tokens strictly from left to right, one step at a time.

The autoregressive core, however, carries structural costs that are increasingly at odds with how speech tokens are actually organized. First, \emph{decoding is serial}: latency and compute grow linearly with the number of generation steps, and the number of steps grows with the audio duration and the token rate. This tension is sharpest for Residual Vector Quantization (RVQ) tokenizers, which represent each frame as a stack of codebook entries: an AR model must impose an artificial generation order on a grid of tokens that has no intrinsic left-to-right structure along the codebook axis. The field has responded with an ecosystem of workarounds: delayed or interleaved codebook streams~\cite{copet2023musicgen,xiaomi2025mimoaudio,gong2026mosstts,defossez2024moshi}, hierarchical time–depth autoregression, in which a large temporal model is paired with a lightweight within-frame decoder~\cite{defossez2024moshi,gong2026mosstts,liao2026fishs2}, and multi-codebook prediction modules that generate residual-codebook tokens from a shared temporal state~\cite{hu2026qwen3tts}. Each is an increasingly sophisticated patch over the same underlying constraint. Second, AR generation is vulnerable to \emph{exposure bias and error accumulation}: a sampling mistake is frozen into the prefix and propagates, manifesting as skipped words and repetitions that TTS systems routinely dedicate training stages to suppressing~\cite{hu2026qwen3tts,song2025distar}. Third, the fixed generation order precludes \emph{bidirectional refinement}: an AR model cannot revisit an early frame in light of later context, and native infilling-style operations such as speech editing require bespoke mechanisms.

Non-autoregressive (NAR) alternatives have long existed on the periphery of this ecosystem. Continuous-space models based on flow matching~\cite{lipman2023flowmatching,le2023voicebox,eskimez2024e2tts,chen2024f5tts} decode rapidly, but they require explicit duration handling~\cite{minimax2025speech}, do not operate in the discrete token space utilizing modern LLM stacks, and consequently forgo the text-knowledge inheritance and serving infrastructure. Masked generative discrete models such as SoundStorm~\cite{borsos2023soundstorm}, NaturalSpeech~3~\cite{ju2024naturalspeech3}, and MaskGCT~\cite{wang2024maskgct} demonstrated years ago that the RVQ token grid can be decoded in parallel by iterative unmasking, yet these systems were built as task-specific architectures at moderate scale, disconnected from the pretraining recipes, initialization strategies, and tooling of general-purpose language models.

That disconnect is now closing from the text side. Diffusion language models (dLLMs), built on masked discrete diffusion~\cite{austin2021d3pm,lou2024sedd,sahoo2024mdlm,shi2024md4}, have matured from proof-of-concept to a competitive paradigm: LLaDA~\cite{nie2025llada} and Dream~\cite{ye2025dream} match similarly sized AR models at the $\sim$8B scale, commercial systems demonstrate order-of-magnitude decoding speedups~\cite{inception2025mercury,song2025seeddiffusion}, and the paradigm has been scaled to 100B parameters~\cite{antgroup2025llada2}. In parallel, \emph{block diffusion}~\cite{arriola2025bd3lm} interpolates between the two extremes: the model is autoregressive over blocks of tokens, retaining KV caching, variable-length generation, and streaming, while denoising all tokens within a block in parallel. Practical recipes for adapting pretrained AR LLMs into (block-)diffusion decoders~\cite{gong2025diffullama,wu2025fastdllmv2,opengvlab2025sdlm} mean that this entire family can inherit the linguistic knowledge of mature text LLMs rather than learning it from scratch. We argue that TTS is a natural, perhaps the natural, application domain for this machinery: the output is strongly grounded in the input text, which collapses the semantic branching factor that makes parallel decoding difficult for free-form text generation~\cite{stepfun2026stepaudio25}, and the RVQ token grid is precisely the kind of order-free, locally correlated structure that any-order masked prediction handles natively and AR decoding does not.

Early explorations support this view but leave the picture incomplete. On the fully parallel side, LLaDA-TTS~\cite{fan2026lladatts} converts a compact AR text LLM into a masked-diffusion speech model and DiffuSpeech~\cite{lou2026diffuspeech} adapts a text dLLM to speech, both at small data scales, while OmniVoice~\cite{zhu2026omnivoice} scales masked-diffusion TTS to 581K hours with an emphasis on massive language coverage. On the block side, DiSTAR~\cite{song2025distar} interleaves AR drafting with masked-diffusion infilling, and Chatterbox-Flash~\cite{seo2026chatterboxflash} converts a pretrained AR TTS decoder into a block-diffusion decoder by fine-tuning. What is still missing is (i) a streaming block-diffusion TTS backed by \emph{large-scale diffusion pretraining}, rather than by post-hoc conversion of an AR checkpoint; (ii) a \emph{controlled comparison} between fully parallel and block-autoregressive diffusion decoding under matched tokenizers, data, and model capacity, so that the community can directly evaluate this architectural choice; and (iii) evidence that diffusion-based TTS can meet the quality bar of production AR systems on major languages rather than in breadth-oriented settings.

This report presents \textbf{\luna{}}, a family of diffusion-language-model-based TTS systems pretrained on \dataScale{} of speech spanning \langList{}. \luna{} comprises two variants that share a single speech tokenizer, data pipeline, and \modelSize{}-parameter backbone lineage, and differ in their generation paradigm:
\begin{itemize}[leftmargin=2em,itemsep=2pt,topsep=3pt]
    \item \textbf{\lunaNAR{}} is a fully non-autoregressive masked-diffusion model, trained with unrestricted random masking over the full RVQ token grid~\cite{zhu2026omnivoice} and decoded by confidence-based iterative parallel sampling. It generates the entire grid of an utterance in a fixed number of refinement steps; a dedicated token-level duration predictor built on a separate text LLM (\durLLM{}) supplies the target length for multilingual input. Because the model formulates generation as any-order infilling, it naturally supports both zero-shot voice cloning and speech editing within a unified framework, without requiring task-specific auxiliary mechanisms.
    \item \textbf{\lunaBlock{}} is a block-autoregressive model built on the block-diffusion objective~\cite{arriola2025bd3lm}, obtained from \lunaNAR{} by continual training for block-causal streaming. It is causal across blocks, supporting KV caching and incremental synthesis in \blockMs{} audio packets, while decoding all frames and codebooks within a block in parallel.
\end{itemize}
The family is built by \emph{progressive adaptation} from a pretrained AR text LLM (\initLLM{}), first to \lunaNAR{} and then to \lunaBlock{} (\S\ref{sec:arch-adaptation}), so that both variants inherit strong multilingual text competence, which is particularly valuable for handling the morphological complexity and distinctive writing systems of Japanese and Korean. Operating directly on acoustic RVQ tokens, they drive the codec decoder without an intermediate flow-matching stage, and the intra-frame codebook stack is absorbed naturally into parallel denoising. Pretraining concludes with an annealed fine-tuning stage on high-quality and expressively annotated speech, adding controllable emotions and non-verbal vocalizations; RL post-training~\cite{shao2024deepseekmath} then optimizes utterance-level rewards for content correctness and speaker similarity directly over the denoising trajectory (\S\ref{sec:rl-post-training}).

Our main contributions are as follows:
\begin{itemize}[leftmargin=2em,itemsep=2pt,topsep=3pt]
    \item \textbf{A diffusion-LLM TTS family at the production scale.} To our knowledge, \luna{} is the largest pretraining effort for diffusion-based TTS to date (\dataScaleShort{} of speech) and the first to target \langList{}. On Seed-TTS-Eval, \lunaNAR{} achieves the best results on all four metrics among the compared open-source and commercial systems (\seedZhCER{} CER / \seedZhSIM{} SIM on test-zh, \seedEnWER{} WER / \seedEnSIM{} SIM on test-en), and it likewise posts the lowest Mandarin and English error rates in our comparison on the in-the-wild CV3-Eval benchmark (\S\ref{sec:evaluation}).
    \item \textbf{A progressive adaptation recipe yielding two deployment-matched variants.} \luna{} is built entirely by continual training (causal $\to$ bidirectional $\to$ block-causal), with each transition inheriting the text competence and speech knowledge already acquired; the same recipe applies to any existing codec language model. To our knowledge, \lunaBlock{} is the first streaming block-diffusion TTS backed by large-scale diffusion pretraining rather than by conversion of an AR checkpoint~\cite{seo2026chatterboxflash}, and because it is derived from \lunaNAR{} under an identical tokenizer, data pipeline, and backbone, the two variants form a controlled realization of the AR--diffusion interpolation~\cite{arriola2025bd3lm}: fully parallel decoding for offline throughput, and block-autoregressive decoding for incremental delivery. Choosing between them is a deployment decision, not a system migration; under the warmed local serving protocol, \lunaBlock{} reaches an end-to-end RTF of 0.0240 and 41.6\,ms local first-block latency for a \blockMs{} block (\S\ref{sec:inference-protocol}).
    \item \textbf{Controllable expressive speech generation.} We extend \lunaNAR{} with explicit control over utterance-level emotion and inline non-verbal vocalizations (NVVs), shaping an utterance as a coherent vocal performance: the intended emotion guides overall delivery, while context-appropriate NVVs are realized at designated positions. Against leading commercial systems, \lunaNAR{} achieves the best results on most objective, model-based, and human-rated metrics for both NVV and emotion control (\S\ref{sec:eval-control}).
    \item \textbf{RL post-training for masked speech-token diffusion.} We show that GRPO-style RL post-training~\cite{shao2024deepseekmath}, previously applied to AR TTS~\cite{cui2025glmtts,liao2026fishs2}, transfers to the masked-diffusion setting by defining policy ratios over the token decisions actually realized along the iterative denoising trajectory, with rollout and replay evaluated under the same effective sampling policy, rather than over a left-to-right factorization (\S\ref{sec:rl-post-training}). The ratios are optimized against a group-relative, lexicographically ranked reward that prioritizes content correctness and breaks ties by speaker similarity.
\end{itemize}

The remainder of this report is organized as follows. \S\ref{sec:architecture} describes the architecture: the \lunaTok{} tokenizer, the shared masked-diffusion formulation over the token grid, and the two decoding regimes built on it. \S\ref{sec:pretraining} details the data and the multi-stage pretraining schedule, the diffusion to block-diffusion adaptation, and the token-level duration predictor. \S\ref{sec:rl-post-training} presents RL post-training over realized denoising trajectories. \S\ref{sec:inference} defines the execution paths, streaming contract, and measurement protocol of the two variants. \S\ref{sec:evaluation} evaluates zero-shot synthesis, expressive control, and dedicated-voice quality, and \S\ref{sec:conclusion} concludes.

\section{Architecture}
\label{sec:architecture}

\subsection{Overview}
\label{sec:arch-overview}

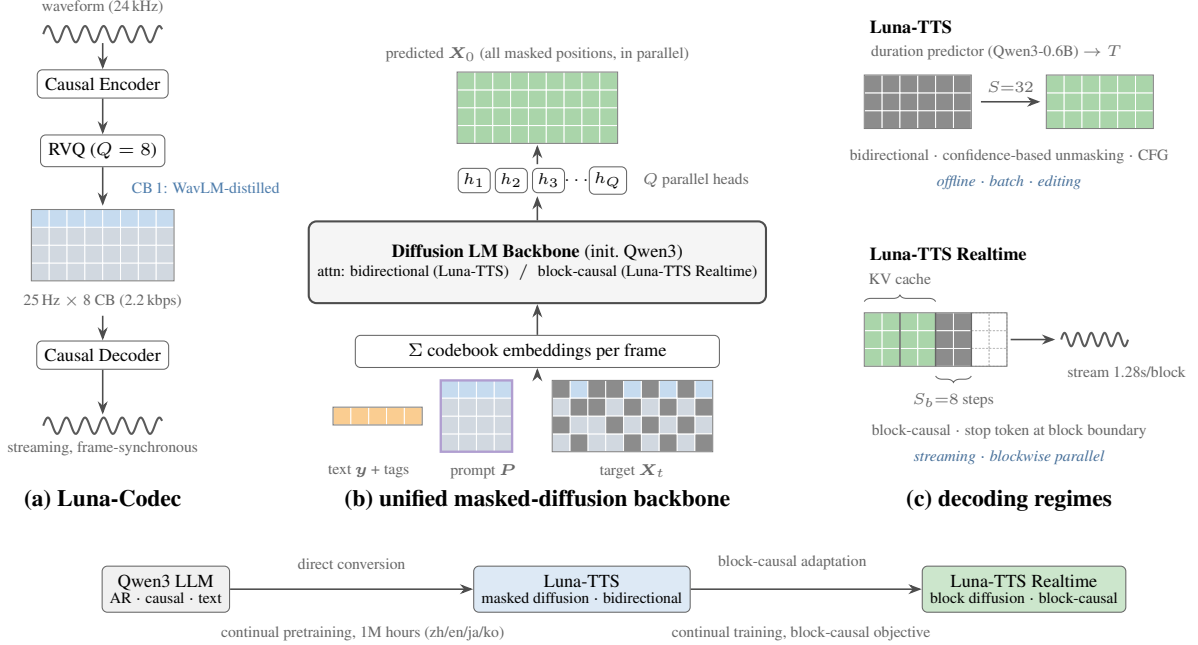
\begin{figure}[t]
    \centering
    \resizebox{\textwidth}{!}{\definecolor{lunaBlue}{RGB}{86,138,190}
\definecolor{lunaBlueL}{RGB}{198,219,239}
\definecolor{lunaGray}{RGB}{209,216,224}
\definecolor{lunaGreen}{RGB}{170,213,170}
\definecolor{lunaOrange}{RGB}{250,205,145}
\definecolor{lunaPurple}{RGB}{176,158,208}
\definecolor{lunaDark}{RGB}{90,90,90}

\begin{tikzpicture}[x=1cm,y=1cm,
  box/.style={draw=black!60, rounded corners=2pt, align=center, font=\scriptsize, inner sep=3pt, fill=white},
  bigbox/.style={draw=black!70, thick, rounded corners=3pt, align=center, font=\scriptsize, fill=black!4},
  plabel/.style={font=\footnotesize\bfseries, align=center},
  sublabel/.style={font=\scriptsize\bfseries, align=center},
  note/.style={font=\tiny, align=center, text=black!60},
  tag/.style={font=\tiny\itshape, align=center, text=lunaBlue!80!black},
  arr/.style={-{Stealth[length=1.8mm]}, semithick, black!70},
  cell/.style={draw=white, line width=0.3pt}]

\def\cs{0.24}

\begin{scope}[shift={(0,0)}]
  \draw[thick, lunaDark] plot[domain=0:1.6, samples=80, smooth] (\x+0.55, {6.62+0.11*sin(\x*1440)});
  \node[note] at (1.35,6.98) {waveform (\sampleRate)};
  \node[box] (enc) at (1.35,5.95) {Causal Encoder};
  \draw[arr] (1.35,6.5) -- (enc.north);
  \node[box] (rvq) at (1.35,5.05) {RVQ ($Q=\numCodebooks$)};
  \draw[arr] (enc.south) -- (rvq.north);
  \begin{scope}[shift={(0.39,3.3)}]
    \foreach \c in {0,...,7} {
      \fill[cell, fill=lunaBlueL] (\c*\cs, 3*\cs) rectangle ++(\cs,\cs);
      \foreach \r in {0,1,2} \fill[cell, fill=lunaGray] (\c*\cs, \r*\cs) rectangle ++(\cs,\cs);
    }
    \draw[black!50, line width=0.4pt] (0,0) rectangle (8*\cs, 4*\cs);
  \end{scope}
  \draw[arr] (rvq.south) -- (1.35,4.31);
  \node[note] at (1.35,3.03) {\frameRate{} $\times$ \numCodebooks{} CB (\bitrate)};
  \node[note, anchor=west, text=lunaBlue!90!black] at (1.62,4.55) {CB\,1: WavLM-distilled};
  \node[box] (dec) at (1.35,2.3) {Causal Decoder};
  \draw[arr] (1.35,2.83) -- (dec.north);
  \draw[thick, lunaDark] plot[domain=0:1.6, samples=80, smooth] (\x+0.55, {1.35+0.11*sin(\x*1440)});
  \draw[arr] (dec.south) -- (1.35,1.5);
  \node[note] at (1.35,1.05) {streaming, frame-synchronous};
  \node[plabel] at (1.35,0.35) {(a) \lunaTok{}};
\end{scope}

\begin{scope}[shift={(4.35,0)}]
  \def\iy{1.0}
  \begin{scope}[shift={(0.10,\iy+0.36)}]
    \foreach \c in {0,...,4} \fill[cell, fill=lunaOrange] (\c*\cs,0) rectangle ++(\cs,\cs);
    \draw[black!50, line width=0.4pt] (0,0) rectangle (5*\cs,\cs);
  \end{scope}
  \node[note] at (0.58,\iy-0.26) {text $\bm{y}$ + tags};
  \begin{scope}[shift={(1.55,\iy)}]
    \foreach \c in {0,...,3} {
      \fill[cell, fill=lunaBlueL] (\c*\cs,3*\cs) rectangle ++(\cs,\cs);
      \foreach \r in {0,1,2} \fill[cell, fill=lunaGray] (\c*\cs,\r*\cs) rectangle ++(\cs,\cs);
    }
    \draw[lunaPurple, line width=0.9pt] (0,0) rectangle (4*\cs,4*\cs);
  \end{scope}
  \node[note] at (2.12,\iy-0.26) {prompt $\bm{P}$};
  \begin{scope}[shift={(3.05,\iy)}]
    \foreach \c in {0,...,8} {
      \fill[cell, fill=lunaBlueL] (\c*\cs,3*\cs) rectangle ++(\cs,\cs);
      \foreach \r in {0,1,2} \fill[cell, fill=lunaGray] (\c*\cs,\r*\cs) rectangle ++(\cs,\cs);
    }
    \foreach \c/\r in {0/3,0/1,1/2,1/0,2/3,2/2,3/1,3/3,4/0,4/2,5/3,5/1,6/2,6/0,7/3,7/1,8/2,8/0}
      \fill[cell, fill=black!45] (\c*\cs,\r*\cs) rectangle ++(\cs,\cs);
    \draw[black!50, line width=0.4pt] (0,0) rectangle (9*\cs,4*\cs);
  \end{scope}
  \node[note] at (4.13,\iy-0.26) {target $\bm{X}_t$};
  \node[box, minimum width=4.9cm] (sum) at (2.85,2.35) {$\Sigma$ codebook embeddings per frame};
  \draw[arr] (2.85,2.02) -- (sum.south);
  \node[bigbox, minimum width=5.4cm, minimum height=1.1cm] (bb) at (2.85,3.55)
    {\textbf{Diffusion LM Backbone} (init.\ \initLLMFamily{})\\[-1pt]
     \tiny attn: bidirectional (\lunaNAR) $\;/\;$ block-causal (\lunaBlock)};
  \draw[arr] (sum.north) -- (bb.south);
  \foreach \i/\x in {1/2.0, 2/2.5, 3/3.0} \node[box, inner sep=2pt, font=\tiny] (h\i) at (\x,4.65) {$h_\i$};
  \node[font=\tiny] at (3.4,4.65) {$\cdots$};
  \node[box, inner sep=2pt, font=\tiny] (hQ) at (3.8,4.65) {$h_Q$};
  \node[note, anchor=west] at (4.15,4.65) {$Q$ parallel heads};
  \draw[arr] (bb.north) -- (2.85,4.44);
  \begin{scope}[shift={(1.77,5.15)}]
    \foreach \c in {0,...,8} \foreach \r in {0,...,3}
      \fill[cell, fill=lunaGreen] (\c*\cs,\r*\cs) rectangle ++(\cs,\cs);
    \draw[black!50, line width=0.4pt] (0,0) rectangle (9*\cs,4*\cs);
  \end{scope}
  \draw[arr] (2.85,4.87) -- (2.85,5.13);
  \node[note] at (2.85,6.35) {predicted $\bm{X}_0$ (all masked positions, in parallel)};
  \node[plabel] at (2.85,0.35) {(b) unified masked-diffusion backbone};
\end{scope}

\begin{scope}[shift={(11.55,0)}]
  \node[sublabel, anchor=west] at (0.0,6.72) {\lunaNAR};
  \node[note, anchor=west] at (0.02,6.38) {duration predictor (\durLLM{}) $\rightarrow T$};
  \begin{scope}[shift={(0.05,5.35)}]
    \foreach \c in {0,...,5} \foreach \r in {0,...,2}
      \fill[cell, fill=black!45] (\c*\cs,\r*\cs) rectangle ++(\cs,\cs);
    \draw[black!50, line width=0.4pt] (0,0) rectangle (6*\cs,3*\cs);
  \end{scope}
  \draw[arr] (1.62,5.71) -- (2.42,5.71) node[midway, above=0pt, note] {$S{=}\narSteps$};
  \begin{scope}[shift={(2.5,5.35)}]
    \foreach \c in {0,...,5} \foreach \r in {0,...,2}
      \fill[cell, fill=lunaGreen] (\c*\cs,\r*\cs) rectangle ++(\cs,\cs);
    \draw[black!50, line width=0.4pt] (0,0) rectangle (6*\cs,3*\cs);
  \end{scope}
  \node[note] at (2.0,4.98) {bidirectional $\cdot$ confidence-based unmasking $\cdot$ CFG};
  \node[tag] at (2.0,4.62) {offline $\cdot$ batch $\cdot$ editing};
  \node[sublabel, anchor=west] at (0.0,3.66) {\lunaBlock};
  \begin{scope}[shift={(0.05,2.15)}]
    \foreach \c in {0,...,3} \foreach \r in {0,...,2}
      \fill[cell, fill=lunaGreen] (\c*\cs,\r*\cs) rectangle ++(\cs,\cs);
    \foreach \c in {4,5} \foreach \r in {0,...,2}
      \fill[cell, fill=black!45] (\c*\cs,\r*\cs) rectangle ++(\cs,\cs);
    \foreach \c in {6,7} \foreach \r in {0,...,2}
      \draw[black!25, line width=0.4pt, dash pattern=on 0.7pt off 0.7pt] (\c*\cs,\r*\cs) rectangle ++(\cs,\cs);
    \draw[black!50, line width=0.4pt] (0,0) rectangle (8*\cs,3*\cs);
    \foreach \c in {2,4,6} \draw[black!60, line width=0.6pt] (\c*\cs,0) -- (\c*\cs,3*\cs);
  \end{scope}
  \draw[decorate, decoration={brace, mirror, amplitude=2.5pt}, black!60]
    (0.05+4*\cs,2.02) -- (0.05+6*\cs,2.02) node[midway, below=3pt, note] {$S_b{=}\blockSteps$ steps};
  \draw[decorate, decoration={brace, amplitude=2.5pt}, black!60]
    (0.05,3.02) -- (0.05+4*\cs,3.02) node[midway, above=3pt, note] {KV cache};
  \draw[arr] (2.02,2.51) -- (2.62,2.51);
  \draw[thick, lunaDark] plot[domain=0:0.9, samples=60, smooth] (\x+2.7, {2.51+0.09*sin(\x*2000)});
  \node[note, anchor=west] at (2.66,2.08) {stream \blockMs{}/block};
  \node[note] at (2.0,1.28) {block-causal $\cdot$ stop token at block boundary};
  \node[tag] at (2.0,0.92) {streaming $\cdot$ blockwise parallel};
  \node[plabel] at (2.0,0.35) {(c) decoding regimes};
\end{scope}

\begin{scope}[shift={(0,-0.85)}]
  \node[box, fill=black!6, minimum height=0.62cm] (L0) at (2.2,0) {\initLLMFamily{} LLM\\[-2pt]\tiny AR $\cdot$ causal $\cdot$ text};
  \node[box, fill=lunaBlueL!60, minimum height=0.62cm] (L1) at (7.8,0) {\lunaNAR\\[-2pt]\tiny masked diffusion $\cdot$ bidirectional};
  \node[box, fill=lunaGreen!60, minimum height=0.62cm] (L2) at (13.75,0) {\lunaBlock\\[-2pt]\tiny block diffusion $\cdot$ block-causal};
  \draw[arr] (L0.east) -- (L1.west) node[midway, above=4pt, note] {direct conversion};
  \node[note] at (4.85,-0.60) {continual pretraining, \dataScaleShort{} (zh/en/ja/ko)};
  \draw[arr] (L1.east) -- (L2.west) node[pos=0.44, above=4pt, note] {block-causal adaptation};
  \node[note] at (10.75,-0.60) {continual training, block-causal objective};
\end{scope}

\end{tikzpicture}}
    \caption{Overview of \luna{}. \textbf{(a)}~\lunaTok{}, a causal RVQ codec (\frameRate{}, $Q=\numCodebooks{}$ codebooks) whose first codebook is semantically anchored by WavLM distillation. \textbf{(b)}~The shared diffusion LM backbone: text, prompt token grid, and partially masked target token grid form one sequence; per-frame codebook embeddings are summed at the input, and $Q$ parallel heads predict all masked positions in parallel. \textbf{(c)}~The two decoding regimes: \lunaNAR{} generates the full token grid in $S$ parallel refinement steps with an external duration predictor; \lunaBlock{} decodes block by block with KV caching and per-block denoising, streaming audio as each block commits. \textbf{Bottom:} the progressive adaptation lineage from a pretrained \initLLMFamily{} LLM through \lunaNAR{} to \lunaBlock{}.}
    \label{fig:overview}
\end{figure}

\luna{} follows a two-component design: a Residual Vector Quantization (RVQ) speech tokenizer, \lunaTok{}, that maps waveforms to a compact discrete token grid and back (\S\ref{sec:arch-tokenizer}); and a \modelSize{}-parameter diffusion language model that generates this grid conditioned on text and an optional reference prompt (\S\ref{sec:arch-formulation}--\S\ref{sec:arch-block}). Figure~\ref{fig:overview} illustrates the overall pipeline.

Three principles drive the design:

\paragraph{Information-complete acoustic modeling in a single generative stage.}
Mainstream LLM-based TTS pipelines predict low-bitrate \emph{semantic} tokens and delegate acoustic rendering to a separate flow-matching or diffusion detokenizer~\citep{du2024cosyvoice,cui2025glmtts,anastassiou2024seedtts,xiang2026qwenaudio3tts}. This split stabilizes AR decoding but creates a semantic--acoustic divide: the language model never sees, and can never optimize for, the acoustic detail that determines speaker fidelity and expressiveness~\citep{zhou2025voxcpm}. \luna{} instead models the full multi-codebook token grid directly, so that a single generative stage owns all information and the codec decoder is a deterministic, lightweight map back to audio.

\paragraph{Any-order generation matches the geometry of the token grid.}
An RVQ-coded utterance is a two-dimensional grid, with time along one axis and codebook depth along the other, whose entries are strongly correlated locally but admit no natural total order along the depth axis. AR systems must nonetheless impose one, via delay patterns~\citep{copet2023musicgen}, per-frame depth Transformers~\citep{defossez2024moshi}, or multi-token-prediction heads~\citep{hu2026qwen3tts}. Masked diffusion dissolves the problem: all grid positions are modeled symmetrically under a single any-order denoising objective, and cross-codebook dependencies are resolved over refinement iterations rather than by architectural decree.

\paragraph{One formulation, two operating points.}
The masked-diffusion objective admits a spectrum of decoding regimes indexed by block size~\citep{arriola2025bd3lm}, from fully parallel generation of an entire utterance to token-by-token AR decoding. \luna{} instantiates two deployment-oriented operating points: \lunaNAR{} for full-grid offline synthesis and native editing, and \lunaBlock{} for block-causal incremental delivery. \lunaBlock{} is obtained from \lunaNAR{} by continual training under a block-causal attention pattern (\S\ref{sec:arch-block}), so the two variants share every other design decision, and choosing between them is a matter of matching the operating point to the workload.

\subsection{Speech Tokenizer: \lunaTok{}}
\label{sec:arch-tokenizer}

\lunaTok{} is a causal neural codec that encodes \sampleRate{} speech into a token grid at \frameRate{} with $Q=\numCodebooks{}$ residual codebooks of \codebookSize{} entries each, i.e., \tokPerSec{} tokens per second at an effective bitrate of \bitrate{}. The encoder is a strided causal convolutional network followed by a lightweight causal Transformer bottleneck; the decoder mirrors this structure and reconstructs the waveform frame-synchronously, so that synthesized audio can be emitted with per-frame granularity during streaming.

\paragraph{Semantically anchored first codebook.}
Purely acoustic RVQ tokens front-load fine spectral detail into early codebooks, which makes the token sequence unnecessarily hard to predict from text~\citep{ye2024xcodec}. Following the semantic-distillation line of work~\citep{zhang2024speechtokenizer,defossez2024moshi}, we regularize the first codebook toward linguistic content by distilling frame-level representations from a pretrained WavLM encoder~\citep{chen2022wavlm}, while leaving the residual codebooks free to capture timbre, prosodic nuance, and channel characteristics. The result is a grid whose depth axis is roughly ordered from linguistic to acoustic information, which keeps the grid predictable from text while retaining the full acoustic detail of the signal.

\paragraph{Training.}
\lunaTok{} is trained on \tokDataScale{} of \langList{} speech with the standard codec recipe: reconstruction losses in the time and multi-scale mel domains, VQ commitment losses, and adversarial training with multi-period and multi-resolution STFT discriminators~\citep{kong2020hifigan,lee2023bigvgan}, plus the first-codebook distillation loss. Quantizer dropout provides bitrate scalability. In both design and reconstruction quality, the resulting tokenizer is in line with those of recent LLM-TTS systems such as Higgs Audio~\citep{bosonai2025higgsaudio} and Qwen3-TTS~\citep{hu2026qwen3tts}.

\subsection{Unified Formulation: Masked Diffusion over the Token Grid}
\label{sec:arch-formulation}

\paragraph{Notation.}
Let $\bm{X}_0 \in \mathcal{V}^{T \times Q}$ denote the token grid (i.e., the $T \times Q$ acoustic code matrix) of a target utterance with $T$ frames and $Q$ codebooks, and let $\bm{c} = (\bm{y}, \bm{P})$ denote the conditioning information: the input text $\bm{y}$ (tokenized by the inherited \initLLM{} BPE vocabulary) and an optional acoustic prompt $\bm{P}$, itself a clean token grid, which specifies the target voice. All conditioning tokens and the $T$ target frames are arranged into a single sequence and processed by a shared Transformer backbone. Each target frame occupies one sequence position, where its $Q$ codebook embeddings are retrieved from separate tables and summed into a single frame-level representation.

\paragraph{Forward process.}
We adopt the absorbing-state masked diffusion process in its simplified form~\citep{sahoo2024mdlm,shi2024md4,nie2025llada}. Given a noise level $t \in (0, 1]$, each target position $(i, q)$ is independently replaced by a special \textsc{mask} token with probability $t$:
\begin{equation}
    q_t\!\left(\bm{X}_t \mid \bm{X}_0\right) \;=\; \prod_{i=1}^{T} \prod_{q=1}^{Q}
    \Big[\, t \cdot \delta_{\textsc{m}}\!\left(X_t^{\,i,q}\right) \;+\; (1 - t) \cdot \delta_{X_0^{\,i,q}}\!\left(X_t^{\,i,q}\right) \Big],
\end{equation}
where $\delta$ denotes a point mass and \textsc{m} the mask token. Conditioning tokens are never masked. The model $p_\theta$ is trained to recover all masked positions in parallel, yielding the weighted denoising cross-entropy that upper-bounds the negative log-likelihood~\citep{sahoo2024mdlm}:
\begin{equation}
    \mathcal{L}_{\mathrm{MD}}(\theta) \;=\;
    - \,\mathbb{E}_{t \sim \mathcal{U}(0,1]}\; \mathbb{E}_{\bm{X}_t \sim q_t}
    \left[ \frac{1}{t} \sum_{(i,q)\,:\,X_t^{\,i,q} = \textsc{m}}
    \log p_\theta\!\left( X_0^{\,i,q} \,\middle|\, \bm{X}_t, \bm{c} \right) \right].
    \label{eq:md-loss}
\end{equation}

\paragraph{Frame-wise multi-codebook modeling.}
The backbone consumes one embedding per \emph{frame}: the $Q$ codebook embeddings of a frame are looked up in separate tables and summed, following the standard practice for multi-stream audio LMs~\citep{defossez2024moshi,gong2026mosstts}, which keeps the Transformer sequence length at $T$ rather than $TQ$ and the LM-side rate at \frameRate{}. Symmetrically, the output hidden state of each frame feeds $Q$ parallel classification heads, one per codebook. Masked positions participate in the sum via a learned per-codebook \textsc{mask} embedding, so a frame may be partially masked along the depth axis. Text embeddings, positional encoding, and the text head are inherited from \initLLM{} unchanged; the audio embedding tables and heads are new parameters. The training loss accumulates the per-position cross-entropy over all $Q=\numCodebooks{}$ codebook heads.

\paragraph{Text-based expressive control.}
Expressive control uses the text-side conditioning pathway. An
utterance-level emotion token in the instruction context specifies the
overall emotional expression, while inline non-verbal vocalization (NVV)
tokens in the synthesis text indicate local vocal events, such as laughter
and coughing. These special control tokens require no dedicated style
encoder or control
head~\citep{liao2026fishs2,stepfun2026stepaudio25,xiang2026qwenaudio3tts}
and are not verbalized as lexical content. Inline NVV tokens specify
text-relative event placement rather than exact acoustic timestamps.
Speaker identity is specified separately by the acoustic prompt $\bm{P}$.
Expressive annotation and continual pretraining are described in
\S\ref{sec:pretrain-data} and
\S\ref{sec:pretrain-schedule}, respectively.

\subsection{From AR Text LLM to Speech Diffusion LM}
\label{sec:arch-adaptation}

Rather than being pretrained from scratch, \luna{} is developed through
\emph{progressive adaptation} across three attention regimes: causal
attention in the AR text model, fully bidirectional attention in
\lunaNAR{}, and block-causal attention in \lunaBlock{}. Each stage is
initialized from the preceding one, preserving previously acquired
representations and capabilities.

The starting point is \initLLM{}, a pretrained autoregressive text LLM\@. The motivation is twofold: text competence (grapheme-to-phoneme regularities, named entities, code-switching, and the orthographic complexity of Japanese and Korean) is expensive to acquire from speech-paired data alone; and prior work shows AR-initialized diffusion models train substantially faster than from-scratch counterparts~\citep{gong2025diffullama,ye2025dream,wu2025fastdllmv2}.

The first transition is a \emph{direct conversion} from \initLLM{} to
\lunaNAR{}. We initialize the backbone with the \initLLM{} weights, replace
the causal attention mask with full bidirectional attention over the entire
sequence, including the text, acoustic prompt, and target token grid, and
continue training on speech--text data using the masked-diffusion objective
in Eq.~\eqref{eq:md-loss}. The AR-pretrained representations transfer
effectively to any-order denoising, consistent with prior findings that AR
checkpoints provide strong initializations for masked-diffusion
decoders~\citep{gong2025diffullama,ye2025dream,fan2026lladatts}. The resulting
model is trained to convergence on the full \dataScaleShort{} corpus, yielding
\lunaNAR{} (\S\ref{sec:arch-nar}). The subsequent transition from
\lunaNAR{} to \lunaBlock{} is described in \S\ref{sec:arch-block}.

\subsection{\lunaNAR{}: Fully Parallel Masked-Diffusion Generation}
\label{sec:arch-nar}

\lunaNAR{} applies Eq.~\eqref{eq:md-loss} over the \emph{entire} utterance, with full bidirectional attention across the whole sequence---conditioning and target grid. Training applies unrestricted random masking over the full $T \times Q$ grid, following \citet{zhu2026omnivoice}: the noise level $t$ is sampled per utterance, and every grid position is masked independently, with no structural constraint along either the time or the codebook axis---the model must learn to complete arbitrary partial grids, which is precisely the capability that iterative parallel decoding, prompt-conditioned cloning, and editing all instantiate as special cases. Generation starts from an all-\textsc{mask} grid of $T$ frames and produces the complete utterance in $S=\narSteps{}$ refinement steps.

\paragraph{Voice cloning and speech editing through infilling.}
The input sequence is $[\bm{y};\, \bm{P};\, \bm{X}_t]$: text, clean prompt grid, and partially masked target grid. Because the prompt is simply an unmasked region of the same grid, zero-shot voice cloning is not a special mechanism but the native infilling behavior of the model: the target region is completed so as to be maximally coherent with the visible acoustic context. The same property naturally extends to speech editing: masking an arbitrary span of an existing utterance and denoising it under the original or modified text regenerates the masked region while maintaining coherence with both the past and \emph{future} audio---a capability that AR models do not natively support without bespoke mechanisms~\citep{fan2026lladatts,wang2024maskgct}.

\paragraph{Token-level duration prediction.}
Fully parallel generation requires the frame count $T$ up front. Rule-based length heuristics (character or phone counts scaled by per-language speaking rates) are brittle exactly where \lunaNAR{} operates: the \numLangs{} supported languages differ markedly both in writing system---Chinese uses characters that represent meaningful syllables, Japanese combines kanji with syllabic kana, Korean uses the Hangul alphabet, and English uses the Latin alphabet---and in how text length maps to speech duration. We therefore train a dedicated \emph{token-level duration predictor}: a separate text LLM (\durLLM{}) fine-tuned to predict a duration for every token of the input text, conditioned on the full sentence context (\S\ref{sec:pretrain-duration}). The per-token predictions are summed to obtain $T$, and the token-level formulation exposes durations at sub-utterance granularity, which supports fine-grained rate manipulation and localized editing. At inference, $T$ may also be set or rescaled by the user, yielding continuous speech-rate control as a free byproduct~\citep{wang2024maskgct}.

\paragraph{Iterative parallel decoding.}
Decoding follows the confidence-based unmasking scheme of the MaskGIT family~\citep{chang2022maskgit,borsos2023soundstorm,nie2025llada}: at each step, the model predicts all masked positions in parallel; a fraction of positions given by a cosine schedule is committed---selected by predicted confidence with annealed Gumbel noise---and the remainder is re-masked for the next iteration. Because training used unrestricted random masking, the decoder is free to commit positions in whatever order confidence dictates---across time and codebook depth---rather than in an order fixed by the architecture. We additionally apply classifier-free guidance on the text condition (dropped with probability $0.1$ during training) to sharpen text adherence at low step counts~\citep{zhu2026omnivoice}. This guided, truncated sampling family (top-$k$ truncation, temperature, CFG) is also the effective policy family under which RL post-training computes its importance ratios (\S\ref{sec:rl-post-training}). The step count $S$ thus acts as an inference-time quality--speed dial, adjustable per request without any retraining.

\paragraph{Strengths and limitations.}
The advantage is a fixed sequential refinement budget with fully parallel
execution over the grid, making \lunaNAR{} the preferred engine for offline
dubbing, short-form narration, and data synthesis. Each forward pass still
scales with the target grid, while the full-sequence formulation precludes
streaming and requires a global duration before synthesis begins. These are the
constraints that motivate \lunaBlock{}.

\subsection{\lunaBlock{}: Block-Causal Streaming Generation}
\label{sec:arch-block}

\lunaBlock{} extends masked-diffusion speech generation to streaming
synthesis by introducing causality only at block boundaries. Rather than
returning to token-level autoregressive decoding, it generates a sequence of
short acoustic blocks while retaining parallel refinement within each block.
Once finalized, a block becomes immutable and can be decoded and delivered
immediately. This design preserves much of the parallelism of \lunaNAR{} while
providing the incremental availability required by streaming applications.

\subsubsection{Block-Causal Design}

Let the target acoustic-token grid be partitioned along time into
$N=\lceil T/B\rceil$ blocks,
$\bm{X}=(\bm{X}^{(1)},\ldots,\bm{X}^{(N)})$. We model the sequence as
\begin{equation}
    p_\theta(\bm{X}\mid\bm{c})
    =
    \prod_{b=1}^{N}
    p_\theta\!\left(
        \bm{X}^{(b)}
        \mid
        \bm{X}^{(<b)},\bm{c}
    \right),
    \label{eq:block-factorization}
\end{equation}
where $\bm{c}$ denotes the linguistic and acoustic conditioning context.
The factorization imposes an ordering over blocks but not over individual
frames or codebooks within a block. Consequently, the active block is refined
as a joint acoustic structure rather than emitted as a token sequence.

We initialize \lunaBlock{} from \lunaNAR{} and adapt it using the
block-diffusion formulation of \citet{arriola2025bd3lm}. During training, the
model predicts a corrupted current block conditioned on the information that
would already be available in a streaming execution. This aligns the learned
conditional distribution with the block-causal generation process while
retaining the masked acoustic-token objective of the non-autoregressive model.
Block causality is therefore part of the learned generation behavior, rather
than an inference-time masking modification alone.

\subsubsection{Streaming Execution}

Streaming generation maintains a simple state invariant: previously completed
blocks form immutable context, whereas only the current block remains subject
to refinement. The conditioning context is established once, after which the
model repeatedly initializes, refines, and commits one acoustic block. Each
committed block is passed to the causal codec for waveform reconstruction and
can be emitted while generation proceeds to the next block.

This separation between committed and mutable state has two consequences.
First, the initial audio becomes available after processing only one block,
rather than after constructing the full utterance. Second, subsequent blocks
can exploit the accumulated acoustic history without revising content that has
already been delivered. The method thus combines causal progression at the
streaming interface with parallel prediction inside each generation unit.

Generation terminates through a learned end-of-speech decision, with an
explicit maximum duration retained as a serving constraint. The primary
operating parameters are the temporal extent of a block and the computation
allocated to its refinement. Smaller blocks improve delivery granularity,
whereas greater refinement can improve the quality of a block before it is
committed. We characterize these latency--quality operating points using the
serving protocol in \S\ref{sec:inference-protocol}.

\FloatBarrier
\subsection{Design Trade-offs at a Glance}
\label{sec:arch-tradeoffs}

Table~\ref{tab:paradigm} summarizes the three-way comparison that \luna{} makes explicit. We regard neither variant as dominant: they are two operating points of one model family, exposed because production workloads genuinely bifurcate into throughput-bound offline synthesis and latency-bound interactive synthesis.

\begin{table}[!ht]
    \centering
    \small
    \caption{Generation-paradigm comparison for a $T$-frame, $Q$-codebook utterance. NFE = number of function evaluations (sequential forward passes). AR figures assume a flattened or delay-pattern decoder over the same grid.}
    \label{tab:paradigm}
    \begin{tabular}{@{}lccc@{}}
        \toprule
        & AR codec LM & \lunaNAR{} & \lunaBlock{} \\
        \midrule
        Generation order & fixed, left-to-right & any-order, global & any-order within block, left-to-right between blocks \\
        Sequential NFE & $\mathcal{O}(T)$--$\mathcal{O}(TQ)$ & $S$ (const.) & $S_b \lceil T/B \rceil$ \\
        KV cache & \checkmark & --- & \checkmark \\
        Streaming & \checkmark & --- & \checkmark (\blockMs{} blocks) \\
        Length handling & stop token & duration predictor & stop token \\
        Intra-frame codebooks & delay / depth-AR / MTP & native parallel & native parallel \\
        Infilling \& editing & bespoke mechanisms & native & prefix-constrained \\
        Error propagation & unbounded prefix & iterative refinement & within-block refinement; irrevocable after commit \\
        Preferred workload & --- & offline / batch & streaming / interactive \\
        \bottomrule
    \end{tabular}
\end{table}
\FloatBarrier

\section{Data and Pretraining}
\label{sec:pretraining}

This section describes the construction of the multilingual training
corpus---including its expressively annotated subset---and the progressive
training schedule that produces \lunaNAR{} and \lunaBlock{}.

\subsection{Data Construction}
\label{sec:pretrain-data}

We construct the training corpus in two parts: a broad multilingual
pretraining mixture, and an expressively annotated subset that carries
utterance-level emotion labels and inline non-verbal vocalization (NVV) tags.

\subsubsection*{Pretraining Data Construction}

We organize data production into four conceptual stages
(Figure~\ref{fig:data-pipeline}):
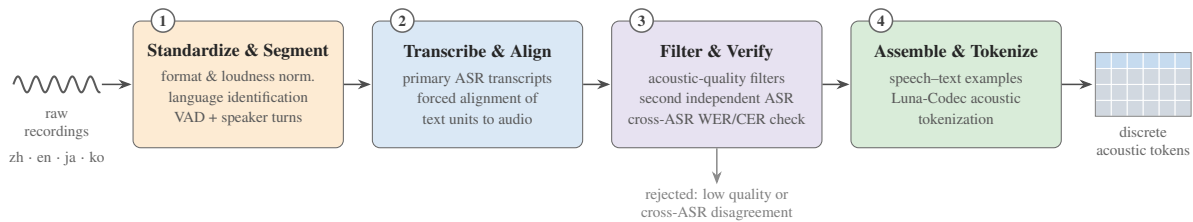
\begin{figure}[!ht]
    \centering
    \resizebox{\linewidth}{!}{\definecolor{lunaBlue}{RGB}{86,138,190}
\definecolor{lunaBlueL}{RGB}{198,219,239}
\definecolor{lunaGray}{RGB}{209,216,224}
\definecolor{lunaGreen}{RGB}{170,213,170}
\definecolor{lunaOrange}{RGB}{250,205,145}
\definecolor{lunaPurple}{RGB}{176,158,208}
\definecolor{lunaDark}{RGB}{90,90,90}

\begin{tikzpicture}[x=1cm,y=1cm,
  stage/.style={draw=black!60, rounded corners=3pt, align=center,
    minimum width=2.9cm, minimum height=1.75cm, inner sep=4pt,
    font=\tiny, text=black!65},
  badge/.style={circle, draw=black!60, fill=white, semithick, inner sep=0pt,
    minimum size=0.32cm, font=\tiny\bfseries, text=black!75},
  note/.style={font=\tiny, align=center, text=black!60},
  arr/.style={-{Stealth[length=1.8mm]}, semithick, black!70},
  cell/.style={draw=white, line width=0.3pt}]

\newcommand{\stagetitle}[1]{{\scriptsize\bfseries\color{black!85}#1}}

\def\cs{0.22}

\draw[thick, lunaDark] plot[domain=0:1.2, samples=70, smooth]
  (\x+0.1, {0.11*sin(\x*1440)});
\node[note] at (0.7,-0.56) {raw\\recordings};
\node[note] at (0.7,-1.02) {zh $\cdot$ en $\cdot$ ja $\cdot$ ko};
\draw[arr] (1.32,0) -- (1.73,0);

\node[stage, fill=lunaOrange!40] (s1) at (3.20,0)
  {\stagetitle{Standardize \& Segment}\\[3pt]
   format \& loudness norm.\\[1pt]
   language identification\\[1pt]
   VAD + speaker turns};
\node[badge] at ([xshift=0.42cm]s1.north west) {1};

\draw[arr] (s1.east) -- (5.03,0);

\node[stage, fill=lunaBlueL!65] (s2) at (6.50,0)
  {\stagetitle{Transcribe \& Align}\\[3pt]
   primary ASR transcripts\\[1pt]
   forced alignment of\\[1pt]
   text units to audio};
\node[badge] at ([xshift=0.42cm]s2.north west) {2};

\draw[arr] (s2.east) -- (8.33,0);

\node[stage, fill=lunaPurple!30] (s3) at (9.80,0)
  {\stagetitle{Filter \& Verify}\\[3pt]
   acoustic-quality filters\\[1pt]
   second independent ASR\\[1pt]
   cross-ASR WER/CER check};
\node[badge] at ([xshift=0.42cm]s3.north west) {3};

\draw[arr] (s3.east) -- (11.63,0);

\draw[arr, black!45] (s3.south) -- (9.80,-1.38);
\node[note, text=black!50] at (9.80,-1.66)
  {rejected: low quality or\\cross-ASR disagreement};

\node[stage, fill=lunaGreen!45] (s4) at (13.10,0)
  {\stagetitle{Assemble \& Tokenize}\\[3pt]
   speech--text examples\\[1pt]
   \lunaTok{} acoustic\\[1pt]
   tokenization};
\node[badge] at ([xshift=0.42cm]s4.north west) {4};

\draw[arr] (s4.east) -- (14.98,0);

\begin{scope}[shift={(15.02,-0.44)}]
  \foreach \c in {0,...,5} {
    \fill[cell, fill=lunaBlueL] (\c*\cs, 3*\cs) rectangle ++(\cs,\cs);
    \foreach \r in {0,1,2} \fill[cell, fill=lunaGray] (\c*\cs, \r*\cs) rectangle ++(\cs,\cs);
  }
  \draw[black!50, line width=0.4pt] (0,0) rectangle (6*\cs, 4*\cs);
\end{scope}
\node[note] at (15.68,-0.78) {discrete\\acoustic tokens};

\end{tikzpicture}}
    \caption{Four-stage multilingual data-processing pipeline. Raw
    recordings are converted into speaker-consistent, transcript-verified
    cuts before acoustic tokenization by \lunaTok{}; cuts that fail the
    quality or cross-ASR consistency checks are discarded.}
    \label{fig:data-pipeline}
\end{figure}

Source recordings are first standardized, language identified, and
segmented using speech activity and speaker boundaries. A primary ASR
system supplies transcripts, while forced alignment maps textual units to
the waveform. Acoustic-quality filtering and a second, independently
trained ASR system then verify each bounded, speaker-consistent cut;
cross-ASR agreement is measured at the word level for English and the
character level for Chinese, Japanese, and Korean. Accepted
speaker-consistent cuts are encoded by \lunaTok{} into the $T\times Q$
RVQ token grid.

This pipeline yields approximately \dataScale{} of curated speech across \langList{}. The corpus fills three roles across the training pipeline (\S\ref{sec:pretrain-schedule}): the broad multilingual mixture used for foundation pretraining, a high-quality subset used for annealing, and the expressive subset described below. These subsets may overlap at the source-recording level but serve different purposes: broad coverage establishes linguistic and speaker diversity, while high-quality and expressive data improve fidelity and controllability. Table~\ref{tab:language-mixture} gives the language composition: an evenly balanced Mandarin--English core, complemented by a combined 13.6\% share of Japanese and Korean.

\begin{table}[!ht]
    \centering
    \small
    \caption{Language composition of the training corpus. Shares are independently rounded.}
    \label{tab:language-mixture}
    \begin{tabular}{@{}lc@{}}
        \toprule
        Language & Share \\
        \midrule
        Mandarin Chinese & 43.4\% \\
        English          & 43.1\% \\
        Japanese         &  6.7\% \\
        Korean           &  6.9\% \\
        \bottomrule
    \end{tabular}
\end{table}

\begingroup
\def\RenderExpressiveData{}
\ifdefined\RenderExpressiveEvaluation

\subsection{Expressive Speech Control}
\label{sec:eval-control}

Our evaluation focuses on two complementary aspects of expressive speech
generation: emotion control and NVV generation. For
both aspects, we assess control accuracy and perceptual quality by benchmarking
\lunaNAR{} against three commercial systems:
MiniMax Speech 2.8 HD~\cite{minimax2026speech28}, ElevenLabs Eleven v3~\cite{elevenlabs2026elevenv3},
and Fish Audio S2.1 Pro~\cite{fishaudio2026s21}.

\subsubsection{Non-Verbal Vocalization Control}

We evaluate NVV control using the Multilingual-NVASR protocol from
NV-Bench~\cite{ni2026nvbench} and the LALM-based perceptual evaluation
protocol from NVV-SuperBench~\cite{xue2026nvvsuperbench}. The objective
evaluation uses samples from both benchmarks, restricted to four categories
recognized by Multilingual-NVASR and supported by all systems---breath,
cough, laughter, and sigh. The perceptual evaluation covers all NVV-SuperBench categories supported by each system.

\textbf{Objective controllability.}
We
compute PCER (CER calculated on extracted NV symbols)
and report micro-averaged event precision, recall, and F1 to characterize prediction correctness and target-event coverage. All metrics are percentages.

\begin{table}[!htbp]
\centering
\caption{NVASR-based objective evaluation of NVV controllability.}
\label{tab:nvv-objective}
\small
\renewcommand{\arraystretch}{1.15}
\setlength{\tabcolsep}{6pt}
\begin{tabular}{lcccc}
\toprule
\textbf{Model}
& \textbf{PCER} $\downarrow$
& \textbf{Precision} $\uparrow$
& \textbf{Recall} $\uparrow$
& \textbf{F1} $\uparrow$ \\
\midrule
MiniMax Speech 2.8 HD & 41.29\% & 78.00\% & 60.50\% & 68.14\% \\
ElevenLabs Eleven v3            & 41.73\% & \textbf{87.45\%} & 59.79\% & 71.02\% \\
Fish Audio S2.1 Pro  & 47.36\% & 77.36\% & 55.59\% & 64.69\% \\
\midrule
\lunaNAR{}              & \textbf{39.95\%} & 79.38\% & \textbf{66.76\%} & \textbf{72.52\%} \\
\bottomrule
\end{tabular}
\end{table}

\textbf{LALM-based perceptual evaluation.}
Gemini 3.1 Pro Preview~\cite{google2026gemini31pro} rates NVV Accuracy, NVV Perceptual Effect (NVV PE), and overall naturalness, quality, and expression. NVV Accuracy measures event realization and positioning, whereas NVV PE captures the perceptual effectiveness of the generated NVV\@. Overall metrics use a 1--5 scale, while NVV-specific metrics use a 0--5 scale (0 = absent or nearly inaudible).

\begin{table}[!htbp]
\centering
\caption{Average LALM-based NVV-control ratings over each system's supported categories.}
\label{tab:nvv-llm}
\small
\renewcommand{\arraystretch}{1.15}
\setlength{\tabcolsep}{2.5pt}
\begin{tabular}{lccccc}
\toprule
\textbf{Model}
& \makecell{\textbf{NVV}\\\textbf{Accuracy} $\uparrow$}
& \makecell{\textbf{NVV}\\\textbf{PE}$\uparrow$}
& \makecell{\textbf{Overall}\\\textbf{Naturalness} $\uparrow$}
& \makecell{\textbf{Overall}\\\textbf{Quality} $\uparrow$}
& \makecell{\textbf{Overall}\\\textbf{Expression} $\uparrow$} \\
\midrule
MiniMax Speech 2.8 HD & \textbf{3.95} & 3.62 & 3.50 & 4.12 & 3.54 \\
ElevenLabs Eleven v3  & 3.65 & \textbf{3.75} & 3.40 & 4.15 & 3.30 \\
Fish Audio S2.1 Pro   & 3.56 & 2.66 & 2.83 & 3.73 & 3.00 \\
\midrule
\lunaNAR{}            & 3.60 & 3.63 & \textbf{3.65} & \textbf{4.18} & \textbf{3.75} \\
\bottomrule
\end{tabular}
\end{table}

\FloatBarrier

\subsubsection{Emotion Control}

We evaluate emotion control using E-Sim, LALM-based ratings, and human MOS\@.
E-Sim uses the ESD test set~\cite{zhou2022emotional}, which provides test texts
and corresponding human reference recordings. For the LALM and human evaluations, we construct a held-out bilingual emotion
test set using LLM-assisted generation. The resulting natural conversational
utterances cover diverse scenarios and undergo quality checks and human review. E-Sim is evaluated on five shared
emotions---Angry, Happy, Sad, Surprise, and Neutral. In contrast, the LALM and
human evaluations cover all emotion categories explicitly supported by each system.

\textbf{Emotion similarity.}
Using \texttt{emotion2vec\_plus\_large}~\cite{ma2024emotion2vec},
we compute E-Sim as the cosine similarity between utterance-level embeddings of generated speech
and paired human recordings.
Because E-Sim may also capture non-emotional acoustic
factors, we interpret it as a complementary metric.

\begin{table}[!htbp]
\centering
\caption{E-Sim ($\uparrow$) across five shared target emotions.}
\label{tab:emotion-objective}
\small
\renewcommand{\arraystretch}{1.15}
\setlength{\tabcolsep}{3pt}
\begin{tabular}{lcccccc}
\toprule
\textbf{Model}
& \textbf{Angry}
& \textbf{Happy}
& \textbf{Sad}
& \textbf{Surprise}
& \textbf{Neutral}
& \textbf{Overall} \\
\midrule
MiniMax Speech 2.8 HD  & 0.530 & \textbf{0.603} & 0.460 & 0.357 & 0.715 & 0.533 \\
ElevenLabs Eleven v3   & 0.544 & 0.513 & 0.481 & \textbf{0.406} & 0.814 & 0.552 \\
Fish Audio S2.1 Pro    & 0.473 & 0.416 & 0.402 & 0.279 & \textbf{0.867} & 0.487 \\
\midrule
\lunaNAR{}             & \textbf{0.608} & 0.497 & \textbf{0.543} & 0.403 & 0.741 & \textbf{0.558} \\
\bottomrule
\end{tabular}
\end{table}

\textbf{LALM-based evaluation.}
Following the model-as-a-judge paradigm of EmergentTTS-Eval~\cite{manku2025emergentttseval} and the distinction between instruction following and perceptual quality in MINT-Bench~\cite{chen2026mintbench}, Gemini 3.1 Pro Preview~\cite{google2026gemini31pro} rates Emotion Match and Expression Quality on a 1--5 scale. These metrics measure target-emotion alignment and the quality of emotional delivery, respectively. 

\textbf{Human listening test.}
In a randomized, blinded listening test, listeners independently rate each
sample on a 1--5 scale for naturalness and human-likeness (N-MOS) and for how
accurately and convincingly the target emotion is conveyed (E-MOS).

\begin{table}[!htbp]
\centering
\caption{Average emotion-control ratings over each system's supported emotions.}
\label{tab:emotion-perceptual}
\label{tab:emotion-lalm}
\label{tab:emotion-human}
\small
\renewcommand{\arraystretch}{1.15}
\setlength{\tabcolsep}{4pt}
\begin{tabular}{@{}lcccc@{}}
\toprule
& \multicolumn{2}{c}{\textbf{LALM-based Evaluation}}
& \multicolumn{2}{c}{\textbf{Human Listening Test}} \\
\cmidrule(lr){2-3}\cmidrule(lr){4-5}
\textbf{Model}
& \makecell{\textbf{Emotion}\\\textbf{Match} $\uparrow$}
& \makecell{\textbf{Expression}\\\textbf{Quality} $\uparrow$}
& \textbf{N-MOS} $\uparrow$
& \textbf{E-MOS} $\uparrow$ \\
\midrule
MiniMax Speech 2.8 HD & \textbf{2.70} & 4.40 & 4.10 & 3.70 \\
ElevenLabs Eleven v3 & 2.67 & 4.49 & \textbf{4.18} & 3.82 \\
Fish Audio S2.1 Pro & 1.93 & 3.94 & 3.95 & 3.55 \\
\midrule
\lunaNAR{} & 2.63 & \textbf{4.54} & 4.14 & \textbf{3.90} \\
\bottomrule
\end{tabular}
\end{table}

\FloatBarrier

\subsubsection{Overall Expressive Control Performance}

Across the NVV and emotion evaluations, \lunaNAR{} delivers the strongest overall balance of controllability, perceptual quality, and expressive coverage among the evaluated systems. For NVV control, it achieves the lowest PCER and the highest recall and F1 score, together with the best LALM-rated overall naturalness, quality, and expression. For emotion control, it achieves the highest overall E-Sim, the best LALM-rated Expression Quality, and the highest human-rated E-MOS, while maintaining competitive Emotion Match and near-best N-MOS across its full supported emotion inventory. Together, these results show that \lunaNAR{} provides broad and effective expressive control without sacrificing perceptual quality or speech naturalness.

\fi

\ifdefined\RenderExpressiveData
\subsubsection*{Expressive Data Construction}

\emph{Data sources.}
The expressive subset is assembled from internal recordings and publicly
available expressive corpora, and spans a broad range of expressive styles:
character and role-play performance, scripted narration and storytelling,
emotional dialogue, and spontaneous natural speech. All recordings undergo
the standard normalization, segmentation, transcription, alignment, and
quality-control pipeline described above before expressive annotation.

\emph{Emotion and NVV annotations.}
We consolidate source-specific labels into a shared control inventory,
retaining categories that are perceptually distinguishable, sufficiently
represented, consistently annotatable, and useful for synthesis control. The
inventory comprises utterance-level emotion labels and inline NVV event labels.
Table~\ref{tab:expressive-taxonomy} provides non-exhaustive examples to
illustrate the annotation format. Synonymous or closely related source labels
are mapped to canonical categories, while ambiguous, inconsistently annotated,
or underrepresented categories are excluded.

\begin{table}[!ht]
    \centering
    \small
    \caption{Illustrative emotion and NVV control annotations.}

    \label{tab:expressive-taxonomy}
    \renewcommand{\arraystretch}{1.12}
    \setlength{\tabcolsep}{5pt}
    \begin{tabular}{@{}
        p{0.14\textwidth}
        p{0.34\textwidth}
        p{0.42\textwidth}@{}}
        \toprule
        \textbf{Annotation type}
        & \textbf{Scope}
        & \textbf{Labels} \\
        \midrule

        Emotion
        & Utterance-level
        & e.g., \texttt{[happy]}, \texttt{[sad]}, \texttt{[angry]},
          \texttt{[fearful]} \\

        NVV
        & Event-level
          
        & e.g., \texttt{[laughs]}, \texttt{[sighs]}, \texttt{[gasps]},
          \texttt{[coughs]} \\

        \bottomrule
    \end{tabular}
\end{table}

\emph{Annotation with multiple models.}
We combine available source labels, recording metadata, and generation
conditions with an annotation pipeline involving Gemini 3.1 Pro Preview and other
large audio-language models. Given the audio and transcript, the models propose
and cross-check utterance-level emotion labels and inline NVV events. Samples
with model disagreement, low-confidence predictions, or complex event
boundaries are routed to human review. We then filter the annotated data for
audio and transcript quality, label reliability, and duplication, using human
spot checks to calibrate and audit the process.

\emph{Final annotation format.}
Annotations are stored in a common schema comprising an utterance-level emotion
field and a text-aligned sequence of NVV events. The resulting corpus contains
emotion-only, NVV-only, jointly annotated, and neutral speech examples. How these annotations are incorporated into model training is described in
\S\ref{sec:arch-formulation}.
\fi

\ifdefined\RenderExpressiveStageThree
\paragraph{Expressive continual pretraining.}
Starting from the checkpoint obtained after high-quality annealing,
training continues on the expressive mixture described above. A small amount
of neutral speech remains in the mixture to help preserve content accuracy,
speaker identity, and natural prosody. The backbone and masked-diffusion
objective remain unchanged. The resulting checkpoint supports emotion control,
inline NVV generation, and their combination, and provides the initialization
for subsequent model adaptation and RL post-training
(\S\ref{sec:rl-post-training}).
\fi

\endgroup

\subsection{Training Stages}
\label{sec:pretrain-schedule}

Training follows a progressive schedule in which each stage inherits the
previous checkpoint while changing the data distribution or the attention
pattern. Table~\ref{tab:training-stages} separates the data, training amount,
and purpose of each stage.
\begin{table}[!ht]
    \centering
    \small
    \setlength{\tabcolsep}{3pt}
    \renewcommand{\arraystretch}{1.08}
    \caption{Progressive training schedule.}
    \label{tab:training-stages}
    \begin{tabular}{@{}>{\raggedright\arraybackslash}p{0.14\textwidth}>{\raggedright\arraybackslash}p{0.25\textwidth}>{\raggedright\arraybackslash}p{0.22\textwidth}>{\raggedright\arraybackslash}p{0.31\textwidth}@{}}
        \toprule
        Stage & Data & Training amount & Purpose \\
        \midrule
        Foundation pretraining
        & Full multilingual mixture
        & ${\sim}$\dataScale{}; ${\sim}1.1$ weighted epochs; ${\sim}100$B packed speech--text tokens
        & Learn multilingual speech generation from \initLLM{} \\
        \addlinespace
        High-quality annealing
        & Curated high-quality speech
        & ${\sim}100$K-hour pool; approximately one epoch
        & Improve fidelity, robustness, and natural prosody \\
        \addlinespace
        Expressive continual pretraining
        & Emotion- and NVV-annotated speech with neutral mixing
        & Compact finely annotated pool
        & Add independent and joint expressive control \\
        \addlinespace
        Block adaptation
        & Multilingual mixture
        & Approximately 20K optimization steps
        & Learn block-causal streaming and termination \\
        \bottomrule
    \end{tabular}
\end{table}
\paragraph{Foundation pretraining and high-quality annealing.}
Foundation pretraining starts from the direct conversion of \initLLM{}
described in \S\ref{sec:arch-adaptation} and uses the masked-diffusion loss
of Eq.~\eqref{eq:md-loss}. Its learning rate follows a cosine schedule from
a peak of $2.1\times10^{-4}$ to $0.1\times$ the peak. A late-stage foundation checkpoint is then further trained on approximately 100K hours of high-quality speech selected from the full training corpus. At the start of this stage, we retain the learning rate specified by the existing schedule rather than restarting it, ensuring a smooth transition when switching datasets. The learning rate is then annealed to $0.1\times$ its peak value over approximately one epoch. Under identical inference settings, this stage
consistently improves the four-language average WER on CV3-Eval over the
branch-point checkpoint.

\begingroup
\def\RenderExpressiveStageThree{}

\endgroup

\paragraph{Token-level duration predictor.}
\label{sec:pretrain-duration}
The duration predictor used by \lunaNAR{} is fine-tuned from \durLLM{} with
full bidirectional attention. It reads the complete input text and predicts
a distribution over quantized durations for every text token in parallel.
Forced alignment supplies the supervision; at inference, the expectation
of each predicted distribution is decoded and the per-token durations are
summed to obtain the target frame count $T$.

\paragraph{Block adaptation.}
\label{sec:pretrain-block}
The converged \lunaNAR{} checkpoint initializes \lunaBlock{}. Global
target bidirectionality is replaced by the two-stream block-causal mask,
and sequence-level corruption by independently sampled per-block noise
levels. The tokenizer, acoustic embedding space, Transformer weights, and
prediction heads are retained. Training and inference use the same
conditional and unconditional block geometry, and the terminal target is
present even when valid speech ends at a block boundary.

\section{Reinforcement Learning for Masked Speech-Token TTS}
\label{sec:rl-post-training}

We apply reinforcement learning (RL) post-training to \lunaNAR{} to optimize
utterance-level attributes that are not directly captured by token-level
supervision, with a particular emphasis on linguistic fidelity and speaker
consistency. Unlike autoregressive models, masked diffusion models refine
multiple speech-token positions in parallel rather than imposing a strict
left-to-right factorization~\citep{sahoo2024mdlm}. Their sampling process must
therefore be represented as an iterative denoising trajectory rather than as a
sequence of conventional next-token decisions.

Prior work on RL for masked discrete diffusion language models (dLLMs)
approximates the intractable output likelihood with a one-step mean-field
surrogate~\citep{zhao2025d1}. In our system, policy updates are instead organized
around the token decisions realized along the denoising trajectory, following
the trajectory-level perspective of diffusion policy optimization~\citep{black2024ddpo}. This construction preserves the native masked-generation
procedure while enabling optimization with non-differentiable speech-level
feedback. We describe below the system-level formulation used for RL
post-training.

\subsection{Trajectory-Aware Policy Optimization}

Following the notation of \S\ref{sec:arch-formulation}, let
\(\bm{c}=(\bm{y},\bm{P})\) denote the conditioning information: the text
input \(\bm{y}\), comprising the reference transcript and the target text,
and the reference acoustic prompt \(\bm{P}\); additional language or style
conditions are omitted for brevity. Given a target length \(T\), the model
generates
\[
  \bm{X}\in\mathcal{V}^{T\times Q},
\]
where \(Q\) is the number of codebooks and \(\mathcal{V}\) is the valid
audio-token vocabulary. Generation induces a sequence of partially completed
states. At denoising step \(k\), the native inference procedure identifies a
set of masked positions \(M_k\) and samples the corresponding token values
\(a_k\), yielding the trajectory
\[
  \tau
  =
  \left\{(s_k,M_k,a_k)\right\}_{k=0}^{S-1},
\]
where \(S\) is the number of refinement steps of the native decoder
(\S\ref{sec:arch-nar}).
The RL update is applied to token decisions conditional on the realized
position sequence. This separation maintains compatibility with the model's
native unmasking procedure without introducing an additional learned position
policy. Joint optimization of the unmasking order would instead require an
explicit stochastic position policy~\citep{raajesh2026maskaware}.

Let \(\pi_\theta(a\mid \bm{c},s,p)\) denote the effective token distribution under
parameters \(\theta\), after applying the same validity constraints and sampling
transformations used by the decoder, including classifier-free guidance when
enabled~\citep{ho2022classifierfree}. Conditioned on the realized set \(M_k\),
the token distribution factorizes as
\[
  \pi_\theta(a_k\mid \bm{c},s_k,M_k)
  =
  \prod_{p\in M_k}\pi_\theta(a_{k,p}\mid \bm{c},s_k,p).
\]
Both rollout and replay evaluate token probabilities within this effective
policy family. This consistency is important because ratios computed from the
untransformed model distribution need not correspond to the distribution that
generated the speech tokens.

Index prompt groups by \(g\), candidate trajectories by \(i\), and committed
token actions by \(j\). On the common valid action domain of the rollout and
replay distributions, the per-action log-ratio and importance ratio are
\[
  \Delta_{g,i,j}(\theta)
  =
  \log \pi_\theta(a_{g,i,j}\mid z_{g,i,j})
  -
  \log \pi_{\mathrm{old}}(a_{g,i,j}\mid z_{g,i,j}),
  \qquad
  r_{g,i,j}(\theta)=\exp\!\left(\Delta_{g,i,j}(\theta)\right),
\]
where \(z_{g,i,j}\) collects the corresponding conditioning, denoising state,
and position. The policy update is thus defined over decisions that were actually
realized during masked generation rather than over an independently
reconstructed terminal sequence.

For each prompt \(g\), the behavior policy samples \(G\geq2\) candidate
utterances and assigns each candidate a scalar utterance-level reward
\(R_{g,i}\). Following group-relative policy optimization (GRPO)~\citep{shao2024deepseekmath}, we compute
\[
  A_{g,i}
  =
  \frac{R_{g,i}-\mu_g}
       {\max(\sigma_g,\sigma_{\min})+\epsilon},
\]
where \(\mu_g\) and \(\sigma_g\) are the mean and sample standard deviation of
the rewards within the prompt group. All token decisions in a trajectory share
the resulting utterance-level advantage.

For a mini-batch \(\mathcal{B}\), the token-level surrogate objective takes the
form
\[
  \mathcal{L}_{\mathrm{RL}}
  =
  \frac{1}{|\mathcal{B}|}
  \sum_{(g,i)\in\mathcal{B}}
  \frac{1}{n_{g,i}}
  \sum_{j=1}^{n_{g,i}}
  \ell_{\mathrm{clip}}\!\left(r_{g,i,j},A_{g,i}\right),
\]
where \(\ell_{\mathrm{clip}}\) is a PPO-style clipped surrogate~\citep{schulman2017ppo} and \(n_{g,i}\) is the number of replayed token
decisions. Applying the surrogate at the token level avoids forming a
trajectory-wide product of ratios across parallel decisions. Per-utterance
normalization further prevents long utterances from receiving disproportionate
weight. Because the objective is additive over recorded actions, replay can be
decomposed over denoising states to control memory consumption without changing
this normalization.

\subsection{Utterance-Level Multi-Objective Feedback}

Each generated waveform is evaluated by frozen models for automatic speech
recognition and speaker verification. Let \(\mathrm{WER}_{g,i}\) denote the word error
rate with respect to the target text and \(\mathrm{SIM}_{g,i}\) the speaker similarity to
the reference audio. Lower values of \(\mathrm{WER}_{g,i}\) indicate higher linguistic
fidelity, whereas higher values of \(\mathrm{SIM}_{g,i}\) indicate stronger speaker
consistency.

The two signals are converted into a prompt-local ordinal reward,
\[
  R_{g,i}
  =
  \Psi_g\!\left(-\mathrm{WER}_{g,i},\,\mathrm{SIM}_{g,i}\right),
\]
where \(\Psi_g\) prioritizes content fidelity and uses speaker similarity to
refine preferences among candidates of comparable linguistic quality. The
policy update is restricted to groups that provide a meaningful relative
preference signal.

This group-local construction avoids two limitations of directly aggregating
raw reward values. First, the scales and reliability of speech metrics can vary
with language, transcript length, prompt difficulty, and reference conditions;
within-prompt comparisons reduce sensitivity to these sources of
heterogeneity. Second, an ordinal composition does not require the numerical
calibration of heterogeneous reward models through a global weighted sum.
The resulting reward therefore expresses the desired task priority while
remaining robust to cross-prompt variation in absolute metric values.

The reward is observed only after the generated token sequence has been decoded
into a waveform. Consequently, the shared trajectory advantage provides coarse
rather than token-specific credit assignment. This design trades fine-grained
credit assignment for stable integration with black-box speech metrics and the
native masked-decoding process.

\section{Inference and Serving}
\label{sec:inference}

\luna{} provides two serving profiles from the same 0.6B backbone.
\lunaNAR{} targets high-throughput offline synthesis, while \lunaBlock{}
targets low-latency streaming synthesis. Both profiles use the same speech
representation and codec interface and are deployed through our vLLM-Omni
serving stack~\citep{yin2026vllmomni}.

\subsection{Serving Profiles}
\label{sec:inference-paths}

\textbf{\lunaNAR{}: full-utterance generation.}
\lunaNAR{} predicts the target duration, initializes the complete acoustic
token canvas, and resolves it through full-grid refinement. The generated
waveform becomes available after all refinement steps and codec decoding
complete. This profile is optimized for offline and batched synthesis.

\textbf{\lunaBlock{}: streaming generation.}
\lunaBlock{} prefills text, inline controls, and reference conditions once,
then generates \blockFrames{}-frame acoustic blocks autoregressively. Each completed block
is decoded immediately and committed to the KV cache. The engine therefore
returns audio while the remaining utterance is still being generated.

Classifier-free guidance can be executed sequentially on one GPU or split
across two GPUs, with the conditional and unconditional branches evaluated in
parallel. The parallel implementation preserves identical acoustic tokens and
waveform bytes.

The engine terminates generation using the learned EOS decision. A configurable
maximum-length bound acts as a serving safeguard.

\subsection{H20 Serving Performance}
\label{sec:inference-latency}
\label{sec:inference-protocol}

We benchmark both profiles in BF16 with batch size 1 on the same
Seed-TTS-Eval Chinese request. Local first-block latency is measured from
engine invocation until the first \blockFrames{}-frame (\blockMs{}) acoustic
block is completed and decoded. Full-response latency includes acoustic
generation, codec decoding, and WAV serialization. All values are medians over
12 warmed runs; network transport is excluded.

\begin{center}
    \centering
    \captionof{table}{\luna{} 0.6B inference performance on NVIDIA H20 GPUs.}
    \label{tab:h20-inference}
    \small
    \setlength{\tabcolsep}{4pt}
    \begin{tabular}{lclrrr}
        \toprule
        Profile & Steps & Execution
            & \makecell{Local first block\\(ms)}
            & \makecell{Full response\\(ms)}
            & \makecell{End-to-end\\RTF $\downarrow$} \\
        \midrule
        \lunaNAR{}   & 32 & $1\times$ H20, full-grid
            & 419.6 & 419.6 & 0.0410 \\
        \lunaNAR{}   & 16 & $1\times$ H20, full-grid
            & 216.0 & \textbf{216.0} & \textbf{0.0211} \\
        \midrule
        \lunaBlock{} & 16 & $1\times$ H20, sequential CFG
            & 98.9 & 837.7 & 0.0790 \\
        \lunaBlock{} & 16 & $2\times$ H20, parallel CFG
            & 63.8 & 451.4 & 0.0426 \\
        \lunaBlock{} & 8 & $1\times$ H20, sequential CFG
            & 59.6 & 457.6 & 0.0432 \\
        \lunaBlock{} & 8 & $2\times$ H20, parallel CFG
            & \textbf{41.6} & 254.0 & 0.0240 \\
        \bottomrule
    \end{tabular}
\end{center}

The two serving profiles optimize different latency targets. At 16 steps,
\lunaNAR{} synthesizes the complete waveform in 216.0\,ms with an end-to-end
RTF of 0.0211. For streaming workloads, \lunaBlock{} returns the first decoded
\blockMs{} audio block in 41.6\,ms and completes a 10.6-second waveform in
254.0\,ms.

Parallel CFG substantially accelerates both operating points. It reduces
full-response latency by 46.1\% at 16 steps and 44.5\% at 8 steps. Local
first-block latency decreases from 98.9 to 63.8\,ms at 16 steps and from
59.6 to 41.6\,ms at 8 steps. We use the 8-step dual-H20 configuration as the
default latency-optimized deployment.

\subsection{Published Latency Context}
\label{sec:inference-context}

Table~\ref{tab:published-rtf} places our measured serving results alongside
published inference results from representative open-source TTS systems. The
configuration and hardware columns retain the operating point reported by each
source.

\begin{center}
    \centering
    \captionof{table}{Inference speed comparison with open-source TTS systems.
    Lower is better.}
    \label{tab:published-rtf}
    \small
    \setlength{\tabcolsep}{5pt}
    \begin{tabular}{llcr}
        \toprule
        System & Configuration & Hardware & RTF $\downarrow$ \\
        \midrule
        \textbf{\lunaNAR{}}
            & 16-step, full-grid & $1\times$ H20 & \textbf{0.0211} \\
        \textbf{\lunaBlock{}}
            & 8-step, parallel CFG & $2\times$ H20 & \textbf{0.0240} \\
        \midrule
        Fast F5-TTS~\citep{zheng2025fastf5latency}
            & 7-step EPSS & RTX 3090 & 0.0300 \\
        OmniVoice~\citep{zhu2026omnivoice}
            & 16-step, batch 1 & H20 & 0.0319 \\
        ZipVoice~\citep{zhu2025zipvoicelatency}
            & 16 NFE & H20 & 0.0557 \\
        VoxCPM2~\citep{openbmb2026voxcpm2latency}
            & accelerated serving & RTX 4090 & $\sim$0.13 \\
        Spark-TTS~\citep{sparkaudio2025sparkttslatency}
            & 0.5B, concurrency 1 & L20 & 0.1362 \\
        F5-TTS~\citep{chen2024f5tts}
            & original model & --- & 0.15 \\
        Fish Audio S2~\citep{liao2026fishs2}
            & SGLang-Omni~\citep{sglangomni2026} & H200 & 0.195 \\
        Qwen3-TTS-25Hz~\citep{hu2026qwen3tts}
            & 0.6B, concurrency 1 & --- & 0.234 \\
        \bottomrule
    \end{tabular}
\end{center}

Among the operating points in Table~\ref{tab:published-rtf}, \lunaNAR{}
achieves the lowest RTF at 0.0211. The streaming \lunaBlock{} profile ranks
second at 0.0240, reducing RTF by 20.0\% relative to Fast F5-TTS and by 24.8\%
relative to OmniVoice while delivering audio block by block.

Table~\ref{tab:published-local-latency} compares published first-output latency
from leading commercial and open-source TTS systems.

\begin{center}
    \centering
    \captionof{table}{Published first-output latency.}
    \label{tab:published-local-latency}
    \small
    \setlength{\tabcolsep}{5pt}
    \begin{tabularx}{0.98\linewidth}{@{}lXr@{}}
        \toprule
        System & Setup & Latency \\
        \midrule
        \textbf{\lunaBlock{}}
            & $2\times$ H20, excluding network
            & \textbf{41.6 ms} \\
        Cartesia Sonic 3.5~\citep{cartesia2026sonic35latency}
            & H100, concurrency 1 & 50 ms \\
        ElevenLabs Flash v2.5~\citep{elevenlabs2026flashv25latency}
            & Excluding application/network & $\sim$75 ms \\
        Hume Octave 2~\citep{hume2026octave2latency}
            & Excluding network & $\sim$100 ms \\
        PlayHT Play 3.0 Mini~\citep{playht2026play30minilatency}
            & --- & 190 ms \\
        Deepgram Aura-2~\citep{deepgram2025aura2latency}
            & --- & $<200$ ms \\
        VibeVoice-Realtime~\citep{microsoft2025vibevoicerealtime}
            & --- & $\sim$300 ms \\
        \bottomrule
    \end{tabularx}
\end{center}

Among the measurements collected in Table~\ref{tab:published-local-latency},
\lunaBlock{} records the lowest first-output latency at 41.6\,ms.

\FloatBarrier
\section{Evaluation}
\label{sec:evaluation}

Our evaluation covers three aspects of the system: zero-shot TTS quality against public benchmarks and strong baselines (\S\ref{sec:eval-zeroshot}), the effectiveness of expressive and paralinguistic control (\S\ref{sec:eval-control}), and the quality of dedicated voices benchmarked against the preset voice libraries of commercial systems (\S\ref{sec:eval-speaker}).

\subsection{Zero-Shot TTS}
\label{sec:eval-zeroshot}

\paragraph{Protocol.}
We evaluate on two public zero-shot benchmarks with full coverage: Seed-TTS-Eval~\citep{anastassiou2024seedtts} (2{,}020 Mandarin / 1{,}088 English utterances; CER via Paraformer-zh for Mandarin, WER via Whisper-large-v3 for English; SIM as cosine similarity of WavLM-large speaker-verification embeddings), and CV3-Eval~\citep{du2025cosyvoice3} scored with the official toolkit (Whisper-large-v3 / Paraformer-zh; SIM via ERes2Net on a 0--100 scale, not comparable to Seed-TTS-Eval SIM). All WER/CER numbers follow the official toolkits' aggregation: the unweighted mean of per-utterance error rates. As strong baselines, we re-evaluate the open-source Qwen3-TTS-12Hz-1.7B-Base~\citep{hu2026qwen3tts} and OmniVoice~\citep{zhu2026omnivoice} (with its official duration estimator) on CV3-Eval, under the same protocol and official scoring toolkits as the \luna{} rows (marked $^\dagger$); numbers for all other systems are quoted from their respective reports or from published comparison tables (marked $^\diamond$) and may differ in evaluation details. The \luna{} rows report the results after RL post-training of \S\ref{sec:rl-post-training}. For \lunaNAR{}, the target length is supplied by the token-level duration predictor of \S\ref{sec:pretrain-duration} with no reference-duration information, and decoding uses $S=32$ refinement steps with classifier-free guidance; \lunaBlock{} is evaluated in its streaming execution mode and instead terminates through its learned end-of-speech decision.

\begin{table}[t]
    \centering
    \small
    \caption{Zero-shot WER/CER (\%) and speaker similarity on Seed-TTS-Eval. $^\diamond$ quoted directly from the original papers or technical reports of the respective models; $^\dagger$ official checkpoint re-evaluated by us using the official evaluation toolkit. SIM is reported on a 100-point scale. Best result per column in bold.}
    \label{tab:seedtts}
    \begin{tabular}{@{}lccccc@{}}
        \toprule
        System & Params & zh CER$\downarrow$ & zh SIM$\uparrow$ & en WER$\downarrow$ & en SIM$\uparrow$ \\
        \midrule
        Seed-TTS$^\diamond$~\citep{anastassiou2024seedtts} & --- & 1.12 & 79.6 & 2.25 & 76.2 \\
        MaskGCT$^\diamond$~\citep{wang2024maskgct} & 1B & 2.27 & 77.4 & 2.62 & 71.7 \\
        F5-TTS$^\diamond$~\citep{chen2024f5tts} & 0.3B & 1.56 & 76.0 & 1.83 & 67.0 \\
        CosyVoice~3$^\diamond$~\citep{du2025cosyvoice3} & 1.5B & 1.12 & 78.1 & 2.21 & 72.0 \\
        MiniMax-Speech$^\diamond$~\citep{minimax2025speech} & --- & 0.83 & 78.3 & 1.65 & 69.2 \\
        GLM-TTS$^\diamond$~\citep{cui2025glmtts} & 1.5B & 1.03 & 76.1 & 2.23 & 67.2 \\
        Qwen3-TTS-12Hz-1.7B-Base$^\dagger$~\citep{hu2026qwen3tts} & 1.7B & 0.98 & 76.9 & 1.68 & 71.7 \\
        Qwen-Audio-3.0-TTS$^\diamond$~\citep{xiang2026qwenaudio3tts} & --- & 0.84 & 79.2 & 1.54 & 76.2 \\
        MOSS-TTS-Local-Transformer$^\diamond$~\citep{gong2026mosstts} & 1.7B & 1.33 & 77.2 & 1.87 & 71.7 \\
        VoxCPM2$^\diamond$~\citep{zhou2026voxcpm2} & 2B & 0.97 & 79.5 & 1.84 & 75.3 \\
        OmniVoice$^\diamond$~\citep{zhu2026omnivoice} & 0.6B & 0.84 & 77.7 & 1.60 & 74.1 \\
        \midrule
        \lunaNAR{} & \modelSize{} & \textbf{\seedZhCER{}} & \textbf{\seedZhSIM{}} & \textbf{\seedEnWER{}} & \textbf{\seedEnSIM{}} \\
        \lunaBlock{} & \modelSize{} & 1.08 & 76.9 & 1.81 & 73.4 \\
        \bottomrule
    \end{tabular}
\end{table}

\begin{table}[t]
    \centering
    \small
    \caption{Zero-shot WER/CER (\%) on the six subsets of CV3-Eval, scored with the official toolkit (CER for zh/ja/ko/hard-zh, WER for en/hard-en). Avg.\ is the mean over the four language subsets (500 utterances each); the hard subsets are reported separately as diagnostics. $^\diamond$ and $^\dagger$ have the same meaning as in Table~\ref{tab:seedtts}; ``---'' indicates values not reported by the source. Best result per column in bold.}
    \label{tab:cv3}
    \begin{tabular}{@{}lccccccc@{}}
        \toprule
        System & zh & en & ja & ko & Avg. & hard-zh & hard-en \\
        \midrule
        CosyVoice~3$^\diamond$~\citep{du2025cosyvoice3} & 3.91 & 4.99 & 7.57 & 5.69 & 5.54 & 9.77 & 10.55 \\
        VoxCPM2$^\diamond$~\citep{zhou2026voxcpm2} & 3.55 & 6.21 & 5.88 & 9.95 & 6.40 & 8.10 & 7.48 \\
        MOSS-TTS-Local-Transformer$^\diamond$~\citep{gong2026mosstts} & 3.95 & 4.35 & 10.10 & 5.95 & 6.09 & --- & --- \\
        Qwen-Audio-3.0-TTS$^\diamond$~\citep{xiang2026qwenaudio3tts} & 3.35 & 4.25 & \textbf{4.78} & \textbf{4.30} & \textbf{4.17} & 7.44 & 6.71 \\
        Qwen3-TTS-12Hz-1.7B-Base$^\dagger$~\citep{hu2026qwen3tts} & 3.19 & 3.92 & 5.00 & 4.63 & 4.18 & 9.36 & 7.47 \\
        OmniVoice$^\dagger$~\citep{zhu2026omnivoice} & 3.89 & 4.57 & 7.24 & 11.80 & 6.88 & 11.98 & 19.69 \\
        \midrule
        \lunaNAR{} & \textbf{\cvZhWER{}} & \textbf{\cvEnWER{}} & \cvJaWER{} & \cvKoWER{} & \cvAvgWER{} & \textbf{\cvHardZhWER{}} & \textbf{\cvHardEnWER{}} \\
        \lunaBlock{} & 3.62 & 4.06 & 6.36 & 5.76 & 4.95 & 12.56 & 13.98 \\
        \bottomrule
    \end{tabular}
\end{table}

\paragraph{Results on Seed-TTS-Eval.}
Table~\ref{tab:seedtts} places \lunaNAR{} first on all four columns. On Mandarin, CER (\seedZhCER{}) leads ahead of MiniMax-Speech (0.83), Qwen-Audio-3.0-TTS and OmniVoice (both 0.84), and the same-protocol Qwen3-TTS baseline (0.98); English WER (\seedEnWER{}) is likewise the lowest, ahead of Qwen-Audio-3.0-TTS (1.54), OmniVoice (1.60), and the same-protocol baseline (1.68). Speaker similarity is the highest on both languages: on English, \seedEnSIM{} leads the next-best entries (76.2 for Seed-TTS and Qwen-Audio-3.0-TTS) by 0.6, while on Mandarin \seedZhSIM{} sits marginally above Seed-TTS (79.6).
\paragraph{Results on CV3-Eval.}
CV3-Eval stresses in-the-wild zero-shot synthesis with substantially harder text---long, irregular, and stylistically diverse---and, unlike Seed-TTS-Eval, covers all four supported languages. \lunaNAR{} posts the lowest English WER in the table (\cvEnWER{}, with the same-protocol Qwen3-TTS baseline at 3.92), is on par with that baseline on Mandarin (\cvZhWER{} vs.\ 3.19), and records the lowest error rates on the in-the-wild hard subsets as well (\cvHardZhWER{} hard-zh and \cvHardEnWER{} hard-en, ahead of Qwen-Audio-3.0-TTS at 7.44 and 6.71); Korean is its hardest subset (\cvKoWER{}). Qwen-Audio-3.0-TTS retains the best Japanese, Korean, and four-language-average results (4.78, 4.30, and 4.17, vs.\ \cvJaWER{}, \cvKoWER{}, and \cvAvgWER{}), with \lunaNAR{} within 0.2 of the best average. Speaker similarity (ERes2Net) averages \cvAvgSIM{} over the six evaluation subsets.

\paragraph{Results for the streaming variant.}
\lunaBlock{} shows that streaming synthesis preserves most of the zero-shot quality: it reaches 1.08 CER / 76.9 SIM on Seed-TTS-Eval test-zh and 1.81 WER / 73.4 SIM on test-en, trading roughly 0.3 CER/WER and 3 SIM points against \lunaNAR{}, and stays close on the CV3-Eval language subsets (average 4.95 vs.\ \cvAvgWER{}). Notably, while operating under streaming constraints, \lunaBlock{} still surpasses several offline-evaluated systems: its CV3-Eval four-language average is better than those of CosyVoice~3 (5.54), MOSS-TTS-Local-Transformer (6.09), and VoxCPM2 (6.40), and its Seed-TTS-Eval Mandarin CER (1.08) is lower than those of Seed-TTS and CosyVoice~3 (both 1.12). The clearest gap is on the CV3-Eval hard subsets (12.56 vs.\ \cvHardZhWER{} on hard-zh, 13.98 vs.\ \cvHardEnWER{} on hard-en), which are demanding on two fronts: the text is long and irregular, and the prompt audio is itself of noticeably lower quality than in the regular subsets. Two properties of block-causal decoding plausibly turn this difficulty into a wider gap. The first is error recovery: difficult text and degraded prompts both raise the chance of flawed early acoustic decisions, which \lunaNAR{} can revise during its global refinement passes, whereas \lunaBlock{} commits each block irrevocably, so an early mistake persists and conditions every subsequent block. The second is length control: \lunaNAR{} receives its frame count from the text-aware duration predictor, whereas \lunaBlock{} must decide when to stop through its learned end-of-speech mechanism, which irregular text and noisy acoustic context can mislead into truncation or run-on errors. Taken together with the serving results of \S\ref{sec:inference}, this quantifies the streaming trade-off: a modest quality cost on regular content and a larger one on hard in-the-wild content, in exchange for blockwise delivery with 41.6\,ms first-block latency.

\begingroup
\def\RenderExpressiveEvaluation{}

\endgroup

\subsection{Dedicated-Voice Evaluation in Our Internal TTS Arena}
\label{sec:eval-speaker}

Beyond zero-shot voice cloning, production TTS systems are often deployed with a fixed roster of dedicated voices. We evaluate this setting using our internal English TTS Arena, comparing \lunaNAR{} with Gemini 3.1 Flash TTS, MiniMax Speech 2.8 HD, ElevenLabs Eleven v3, StepAudio 2.5 TTS, and Inworld TTS 2. To control for differences in voice-library size, three voices are included for each system: our entry uses three dedicated voices obtained by further fine-tuning \lunaNAR{} on internal recordings, while each commercial system uses three voices sampled at random from the preset voices officially recommended on its website. The analysis covers 1{,}319 pairwise comparisons on 100 English prompts spanning customer service, assistants, knowledge sharing, and entertainment.

In every trial, two anonymized samples are presented in randomized left--right order and evaluated for overall speech quality, including naturalness, expressiveness, prosody, and delivery. The system identity, model name, and voice identity are hidden during evaluation. The Arena ranking is estimated using a Bradley--Terry model fitted to the pairwise outcomes, with strengths reported on an Elo-like scale whose mean is anchored at 1500. To assess sampling uncertainty, we performed 300 bootstrap resamples, recomputed the ranking for each resample, and reported the median Elo estimate together with the 5th--95th percentile interval.

As shown in Table~\ref{tab:arena-ranking}, \lunaNAR{} achieves the highest Elo estimate among the six evaluated systems, with a bootstrap-median Elo of 1548.47. It records 495 wins and 368 losses, corresponding to an overall win rate of 57.36\%. Gemini is a close second, and its bootstrap interval overlaps with that of \lunaNAR{}. Therefore, \lunaNAR{} has the highest nominal ranking in this Arena evaluation, although the available evidence does not establish a statistically significant difference between \lunaNAR{} and Gemini 3.1 Flash TTS\@.

\begin{table}[!htbp]
    \centering
    \small
    \caption{Results from our internal English TTS Arena. Elo values are medians from 300 bootstrap resamples, and uncertainty intervals span the 5th--95th bootstrap percentiles.}
    \label{tab:arena-ranking}

    \begin{tabular}{@{}ccccccc@{}}
        \toprule
        Rank & System & Elo & 90\% bootstrap interval & Wins & Losses & Win rate \\
        \midrule
        1 & \textbf{\lunaNAR{}} & \textbf{1548.47} & \textbf{1531.98--1566.59} & \textbf{495} & \textbf{368} & \textbf{57.36\%} \\
        2 & Gemini 3.1 Flash TTS & 1546.13 & 1517.81--1572.46 & 221 & 196 & 53.00\% \\
        3 & Inworld TTS 2 & 1495.67 & 1474.79--1519.24 & 283 & 317 & 47.17\% \\
        4 & ElevenLabs Eleven v3 & 1484.37 & 1436.10--1529.14 & 50 & 69 & 42.02\% \\
        5 & StepAudio 2.5 TTS & 1480.39 & 1458.27--1506.09 & 229 & 298 & 43.45\% \\
        6 & MiniMax Speech 2.8 HD & 1445.53 & 1391.82--1495.23 & 41 & 71 & 36.61\% \\
        \bottomrule
    \end{tabular}
\end{table}

Table~\ref{tab:arena-head2head} reports \lunaNAR{}'s direct results against each competing system. \lunaNAR{} achieves a positive win rate in all five head-to-head comparisons. Its largest margin is observed against Inworld TTS 2, while its comparison with Gemini is nearly balanced.

\begin{table}[!ht]
    \centering
    \small
    \caption{\lunaNAR{} head-to-head results in our internal English TTS Arena.}
    \label{tab:arena-head2head}

    \begin{tabular}{@{}lcc@{}}
        \toprule
        Opponent & \lunaNAR{} wins--losses & Win rate \\
        \midrule
        Inworld TTS 2 & 92--52 & 63.89\% \\
        MiniMax Speech 2.8 HD & 54--34 & 61.36\% \\
        ElevenLabs Eleven v3 & 56--38 & 59.57\% \\
        StepAudio 2.5 TTS & 176--131 & 57.33\% \\
        Gemini 3.1 Flash TTS & 117--113 & 50.87\% \\
        \midrule
        Overall & 495--368 & 57.36\% \\
        \bottomrule
    \end{tabular}
\end{table}

\FloatBarrier
\section{Conclusion}
\label{sec:conclusion}

This report introduced \luna{}, a family of diffusion-LLM-based TTS systems developed through progressive adaptation. A pretrained AR text LLM is first converted into the fully parallel \lunaNAR{} and subsequently adapted into the block-causal \lunaBlock{}. The resulting variants share one model lineage while targeting complementary deployment scenarios: high-throughput offline synthesis and low-latency streaming.

Across quality and controllability, \lunaNAR{} ranks first on all four Seed-TTS-Eval metrics, records the lowest Mandarin and English error rates in our CV3-Eval comparison, and achieves the strongest overall expressive-control results among the evaluated systems. On the systems side, \lunaNAR{} reaches an end-to-end RTF of 0.0211, while \lunaBlock{} combines an RTF of 0.0240 with 41.6\,ms local first-block latency for a \blockMs{} block under the warmed serving protocol. Taken together, these results advance the state of the art in both speech-generation performance and serving efficiency. They demonstrate that masked acoustic-token diffusion can support high-quality offline and real-time synthesis within a unified model family, rather than serving only as an alternative to conventional left-to-right acoustic decoding.

\paragraph{Limitations.}
Among the four supported languages, Korean shows the weakest performance on CV3-Eval, with a CER of \cvKoWER{}. This gap may partly reflect the relatively small proportion of Korean data (6.9\%) in the training mixture. Language coverage is also limited to four languages, substantially narrower than that of massively multilingual TTS systems. In addition, \lunaNAR{} relies on an external duration predictor to determine the target frame count before synthesis, while \lunaBlock{} uses a fixed block size of \blockMs{}, selected for streaming efficiency rather than dynamically adapted to the input content, and trails \lunaNAR{} most visibly on the hard CV3-Eval subsets.

\paragraph{Future work.}
Promising directions for future work include expanding language and dialect coverage, dynamically adapting block sizes to the input content, jointly optimizing the unmasking-position and token policies through reinforcement learning, and integrating duration prediction into the backbone.

\section{Authors}
\label{sec:authors}

\noindent
\textbf{Core Contributors}
\par\vspace{0.6em}

\noindent
\begin{tabularx}{\textwidth}{
  @{}
  *{5}{>{\raggedright\arraybackslash}X}
  @{}
}
Feng Yin\textsuperscript{*}
  & Shuai Shi\textsuperscript{*}
  & Junjie Zheng\textsuperscript{*}
  & Kechenying Zhou\textsuperscript{*}
  & Yiqiu Wang
\\
Chenyang He
  & Qiuhua Jiang
  & Mengxiao Bi
  & Yanmin Qian\textsuperscript{\(\dagger\)}
  &
\end{tabularx}

\vspace{1.2em}

\noindent
\textbf{Contributors\textsuperscript{\(\ddagger\)}}
\par\vspace{0.6em}

\noindent
\begin{tabularx}{\textwidth}{
  @{}
  *{5}{>{\raggedright\arraybackslash}X}
  @{}
}
Mingxin Chen
  & Xun Gong
  & Tianteng Gu
  & Bing Han
  & Peng Jiang
\\
Chenda Li
  & Haiyang Sun
  & Han Wang
  & Wei Wang
  & Yi Wang
\\
Leying Zhang
  & Wangyou Zhang
  & Chushu Zhou
  &
  &
\end{tabularx}

{
\renewcommand{\thefootnote}{}
\footnotetext{%
  \textsuperscript{*} Equal contribution.
  \qquad
  \textsuperscript{\(\dagger\)} Corresponding author.
  \qquad
  \textsuperscript{\(\ddagger\)}
  Listed alphabetically by family name.%
}
}

\bibliographystyle{unsrtnat}
\bibliography{references}

\begin{thebibliography}{86}
\providecommand{\natexlab}[1]{#1}
\providecommand{\url}[1]{\texttt{#1}}
\expandafter\ifx\csname urlstyle\endcsname\relax
  \providecommand{\doi}[1]{doi: #1}\else
  \providecommand{\doi}{doi: \begingroup \urlstyle{rm}\Url}\fi

\bibitem[Zeghidour et~al.(2022)Zeghidour, Luebs, Omran, Skoglund, and
  Tagliasacchi]{zeghidour2021soundstream}
Neil Zeghidour, Alejandro Luebs, Ahmed Omran, Jan Skoglund, and Marco
  Tagliasacchi.
\newblock {SoundStream}: An end-to-end neural audio codec.
\newblock \emph{IEEE/ACM Transactions on Audio, Speech, and Language
  Processing}, 30:\penalty0 495--507, 2022.

\bibitem[D{\'e}fossez et~al.(2022)D{\'e}fossez, Copet, Synnaeve, and
  Adi]{defossez2022encodec}
Alexandre D{\'e}fossez, Jade Copet, Gabriel Synnaeve, and Yossi Adi.
\newblock High fidelity neural audio compression.
\newblock \emph{arXiv preprint arXiv:2210.13438}, 2022.

\bibitem[Kumar et~al.(2023)Kumar, Seetharaman, Luebs, Kumar, and
  Kumar]{kumar2023dac}
Rithesh Kumar, Prem Seetharaman, Alejandro Luebs, Ishaan Kumar, and Kundan
  Kumar.
\newblock High-fidelity audio compression with improved {RVQGAN}.
\newblock In \emph{Advances in Neural Information Processing Systems}, 2023.

\bibitem[Betker(2023)]{betker2023tortoise}
James Betker.
\newblock Better speech synthesis through scaling.
\newblock \emph{arXiv preprint arXiv:2305.07243}, 2023.

\bibitem[Wang et~al.(2023)Wang, Chen, Wu, Zhang, Zhou, Liu, Chen, Liu, Wang,
  Li, He, Zhao, and Wei]{wang2023valle}
Chengyi Wang, Sanyuan Chen, Yu~Wu, Ziqiang Zhang, Long Zhou, Shujie Liu, Zhuo
  Chen, Yanqing Liu, Huaming Wang, Jinyu Li, Lei He, Sheng Zhao, and Furu Wei.
\newblock Neural codec language models are zero-shot text to speech
  synthesizers.
\newblock \emph{arXiv preprint arXiv:2301.02111}, 2023.

\bibitem[Anastassiou et~al.(2024)Anastassiou, Chen, Chen, Chen, Chen, Chen,
  et~al.]{anastassiou2024seedtts}
Philip Anastassiou, Jiawei Chen, Jitong Chen, Yuanzhe Chen, Zhuo Chen, Ziyi
  Chen, et~al.
\newblock Seed-{TTS}: A family of high-quality versatile speech generation
  models.
\newblock \emph{arXiv preprint arXiv:2406.02430}, 2024.

\bibitem[Du et~al.(2024{\natexlab{a}})Du, Chen, Zhang, Hu, Lu, Yang, Hu, Zheng,
  Gu, Ma, Gao, and Yan]{du2024cosyvoice}
Zhihao Du, Qian Chen, Shiliang Zhang, Kai Hu, Heng Lu, Yexin Yang, Hangrui Hu,
  Siqi Zheng, Yue Gu, Ziyang Ma, Zhifu Gao, and Zhijie Yan.
\newblock {CosyVoice}: A scalable multilingual zero-shot text-to-speech
  synthesizer based on supervised semantic tokens.
\newblock \emph{arXiv preprint arXiv:2407.05407}, 2024{\natexlab{a}}.

\bibitem[Du et~al.(2024{\natexlab{b}})Du, Wang, Chen, Shi, Lv, Zhao, Gao, Yang,
  et~al.]{du2024cosyvoice2}
Zhihao Du, Yuxuan Wang, Qian Chen, Xian Shi, Xiang Lv, Tianyu Zhao, Zhifu Gao,
  Yexin Yang, et~al.
\newblock {CosyVoice 2}: Scalable streaming speech synthesis with large
  language models.
\newblock \emph{arXiv preprint arXiv:2412.10117}, 2024{\natexlab{b}}.

\bibitem[Du et~al.(2025)Du, Gao, Wang, Yu, Zhao, Wang, Lv, Wang,
  et~al.]{du2025cosyvoice3}
Zhihao Du, Changfeng Gao, Yuxuan Wang, Fan Yu, Tianyu Zhao, Hao Wang, Xiang Lv,
  Hui Wang, et~al.
\newblock {CosyVoice 3}: Towards in-the-wild speech generation via scaling-up
  and post-training.
\newblock \emph{arXiv preprint arXiv:2505.17589}, 2025.

\bibitem[Zhang et~al.(2025)Zhang, Guo, Yang, Yu, Zhang,
  et~al.]{minimax2025speech}
Bowen Zhang, Congchao Guo, Geng Yang, Hang Yu, Haozhe Zhang, et~al.
\newblock {MiniMax-Speech}: Intrinsic zero-shot text-to-speech with a learnable
  speaker encoder.
\newblock \emph{arXiv preprint arXiv:2505.07916}, 2025.

\bibitem[Ye et~al.(2025{\natexlab{a}})Ye, Zhu, Chan, Wang, Tan, Lei, Peng,
  et~al.]{ye2025llasa}
Zhen Ye, Xinfa Zhu, Chi-Min Chan, Xinsheng Wang, Xu~Tan, Jiahe Lei, Yi~Peng,
  et~al.
\newblock Llasa: Scaling train-time and inference-time compute for
  {Llama}-based speech synthesis.
\newblock \emph{arXiv preprint arXiv:2502.04128}, 2025{\natexlab{a}}.

\bibitem[Cui et~al.(2025)Cui, Yang, Li, Tian, Ma, Zhang, Chen, Yang, Huang,
  Cheng, Zhou, Yu, Gu, and Tang]{cui2025glmtts}
Jiayan Cui, Zhihan Yang, Naihan Li, Jiankun Tian, Xingyu Ma, Yi~Zhang, Guangyu
  Chen, Runxuan Yang, Zijian Huang, Yuqing Cheng, Yizhi Zhou, Guochen Yu,
  Xiaotao Gu, and Jie Tang.
\newblock {GLM-TTS} technical report.
\newblock \emph{arXiv preprint arXiv:2512.14291}, 2025.

\bibitem[Hu et~al.(2026)Hu, Zhu, He, Guo, Zhang, et~al.]{hu2026qwen3tts}
Hangrui Hu, Xinfa Zhu, Ting He, Dake Guo, Bin Zhang, et~al.
\newblock {Qwen3-TTS} technical report.
\newblock \emph{arXiv preprint arXiv:2601.15621}, 2026.

\bibitem[Xiang et~al.(2026)Xiang, Wen, Zhao, Wang, Wang,
  et~al.]{xiang2026qwenaudio3tts}
Bajian Xiang, Cheng Wen, Han Zhao, Hao Wang, Haoxu Wang, et~al.
\newblock {Qwen-Audio-3.0-TTS}: Freely controllable and highly robust speech
  synthesis with multi-stage training paradigm.
\newblock \emph{arXiv preprint arXiv:2607.23938}, 2026.

\bibitem[Liao et~al.(2026)Liao, Wang, Liu, Cheng, et~al.]{liao2026fishs2}
Shijia Liao, Yuxuan Wang, Songting Liu, Yifan Cheng, et~al.
\newblock Fish audio {S2} technical report.
\newblock \emph{arXiv preprint arXiv:2603.08823}, 2026.

\bibitem[{SII-OpenMOSS Team}(2026)]{gong2026mosstts}
{SII-OpenMOSS Team}.
\newblock {MOSS-TTS} technical report.
\newblock \emph{arXiv preprint arXiv:2603.18090}, 2026.

\bibitem[Wang et~al.(2025{\natexlab{a}})Wang, Jiang, Ma, Zhang, Liu, Li,
  et~al.]{wang2025sparktts}
Xinsheng Wang, Mingqi Jiang, Ziyang Ma, Ziyu Zhang, Songxiang Liu, Linqin Li,
  et~al.
\newblock Spark-{TTS}: An efficient {LLM}-based text-to-speech model with
  single-stream decoupled speech tokens.
\newblock \emph{arXiv preprint arXiv:2503.01710}, 2025{\natexlab{a}}.

\bibitem[Zhou et~al.(2025{\natexlab{a}})Zhou, Zhou, He, Zhou, Wang, Deng, and
  Shu]{zhou2025indextts2}
Siyi Zhou, Yiquan Zhou, Yi~He, Xun Zhou, Jinchao Wang, Wei Deng, and Jingchen
  Shu.
\newblock {IndexTTS2}: A breakthrough in emotionally expressive and
  duration-controlled auto-regressive zero-shot text-to-speech.
\newblock \emph{arXiv preprint arXiv:2506.21619}, 2025{\natexlab{a}}.

\bibitem[Guo et~al.(2024)Guo, Liu, Shen, Wu, Xie, Xie, and
  Xu]{guo2024fireredtts}
Hao-Han Guo, Kun Liu, Fei-Yu Shen, Yi-Chen Wu, Feng-Long Xie, Kun Xie, and
  Kai-Tuo Xu.
\newblock {FireRedTTS}: A foundation text-to-speech framework for
  industry-level generative speech applications.
\newblock \emph{arXiv preprint arXiv:2409.03283}, 2024.

\bibitem[{LLM-Core Xiaomi}(2025)]{xiaomi2025mimoaudio}
{LLM-Core Xiaomi}.
\newblock {MiMo-Audio}: Audio language models are few-shot learners.
\newblock \emph{arXiv preprint arXiv:2512.23808}, 2025.

\bibitem[{StepFun Audio Team}(2025)]{stepfun2025stepaudio2}
{StepFun Audio Team}.
\newblock {Step-Audio 2} technical report.
\newblock \emph{arXiv preprint arXiv:2507.16632}, 2025.

\bibitem[Copet et~al.(2023)Copet, Kreuk, Gat, Remez, Kant, Synnaeve, Adi, and
  D{\'e}fossez]{copet2023musicgen}
Jade Copet, Felix Kreuk, Itai Gat, Tal Remez, David Kant, Gabriel Synnaeve,
  Yossi Adi, and Alexandre D{\'e}fossez.
\newblock Simple and controllable music generation.
\newblock In \emph{Advances in Neural Information Processing Systems}, 2023.

\bibitem[D{\'e}fossez et~al.(2024)D{\'e}fossez, Mazar{\'e}, Orsini, Royer,
  P{\'e}rez, J{\'e}gou, Grave, and Zeghidour]{defossez2024moshi}
Alexandre D{\'e}fossez, Laurent Mazar{\'e}, Manu Orsini, Am{\'e}lie Royer,
  Patrick P{\'e}rez, Herv{\'e} J{\'e}gou, Edouard Grave, and Neil Zeghidour.
\newblock Moshi: A speech-text foundation model for real-time dialogue.
\newblock \emph{arXiv preprint arXiv:2410.00037}, 2024.

\bibitem[Song et~al.(2025{\natexlab{a}})Song, Zhuang, Chen,
  et~al.]{song2025distar}
Yakun Song, Xiaobin Zhuang, Jiawei Chen, et~al.
\newblock {DiSTAR}: Diffusion over a scalable token autoregressive
  representation for speech generation.
\newblock \emph{arXiv preprint arXiv:2510.12210}, 2025{\natexlab{a}}.

\bibitem[Lipman et~al.(2023)Lipman, Chen, Ben-Hamu, Nickel, and
  Le]{lipman2023flowmatching}
Yaron Lipman, Ricky T.~Q. Chen, Heli Ben-Hamu, Maximilian Nickel, and Matt Le.
\newblock Flow matching for generative modeling.
\newblock In \emph{International Conference on Learning Representations}, 2023.

\bibitem[Le et~al.(2023)Le, Vyas, Shi, Karrer, Sari, Moritz, Williamson,
  Manohar, Adi, Mahadeokar, and Hsu]{le2023voicebox}
Matthew Le, Apoorv Vyas, Bowen Shi, Brian Karrer, Leda Sari, Rashel Moritz,
  Mary Williamson, Vimal Manohar, Yossi Adi, Jay Mahadeokar, and Wei-Ning Hsu.
\newblock Voicebox: Text-guided multilingual universal speech generation at
  scale.
\newblock In \emph{Advances in Neural Information Processing Systems}, 2023.

\bibitem[Eskimez et~al.(2024)Eskimez, Wang, Thakker, Li, Tsai, Xiao, Yang, Zhu,
  Tang, Tan, Liu, Zhao, and Kanda]{eskimez2024e2tts}
Sefik~Emre Eskimez, Xiaofei Wang, Manthan Thakker, Canrun Li, Chung-Hsien Tsai,
  Zhen Xiao, Hemin Yang, Zirun Zhu, Min Tang, Xu~Tan, Yanqing Liu, Sheng Zhao,
  and Naoyuki Kanda.
\newblock {E2 TTS}: Embarrassingly easy fully non-autoregressive zero-shot
  {TTS}.
\newblock In \emph{IEEE Spoken Language Technology Workshop (SLT)}, 2024.

\bibitem[Chen et~al.(2024)Chen, Niu, Ma, Deng, Wang, Zhao, Yu, and
  Chen]{chen2024f5tts}
Yushen Chen, Zhikang Niu, Ziyang Ma, Keqi Deng, Chunhui Wang, Jian Zhao, Kai
  Yu, and Xie Chen.
\newblock {F5-TTS}: A fairytaler that fakes fluent and faithful speech with
  flow matching.
\newblock \emph{arXiv preprint arXiv:2410.06885}, 2024.

\bibitem[Borsos et~al.(2023)Borsos, Sharifi, Vincent, Kharitonov, Zeghidour,
  and Tagliasacchi]{borsos2023soundstorm}
Zal{\'a}n Borsos, Matt Sharifi, Damien Vincent, Eugene Kharitonov, Neil
  Zeghidour, and Marco Tagliasacchi.
\newblock {SoundStorm}: Efficient parallel audio generation.
\newblock \emph{arXiv preprint arXiv:2305.09636}, 2023.

\bibitem[Ju et~al.(2024)Ju, Wang, Shen, Tan, Xin, Yang, Liu,
  et~al.]{ju2024naturalspeech3}
Zeqian Ju, Yuancheng Wang, Kai Shen, Xu~Tan, Detai Xin, Dongchao Yang, Yanqing
  Liu, et~al.
\newblock {NaturalSpeech 3}: Zero-shot speech synthesis with factorized codec
  and diffusion models.
\newblock In \emph{International Conference on Machine Learning}, 2024.

\bibitem[Wang et~al.(2025{\natexlab{b}})Wang, Zhan, Liu, Zeng, Guo, Zheng,
  Zhang, Zhang, Zhang, and Wu]{wang2024maskgct}
Yuancheng Wang, Haoyue Zhan, Liwei Liu, Ruihong Zeng, Haotian Guo, Jiachen
  Zheng, Qiang Zhang, Xueyao Zhang, Shunsi Zhang, and Zhizheng Wu.
\newblock {MaskGCT}: Zero-shot text-to-speech with masked generative codec
  transformer.
\newblock In \emph{International Conference on Learning Representations},
  2025{\natexlab{b}}.

\bibitem[Austin et~al.(2021)Austin, Johnson, Ho, Tarlow, and van~den
  Berg]{austin2021d3pm}
Jacob Austin, Daniel~D. Johnson, Jonathan Ho, Daniel Tarlow, and Rianne van~den
  Berg.
\newblock Structured denoising diffusion models in discrete state-spaces.
\newblock In \emph{Advances in Neural Information Processing Systems}, 2021.

\bibitem[Lou et~al.(2024)Lou, Meng, and Ermon]{lou2024sedd}
Aaron Lou, Chenlin Meng, and Stefano Ermon.
\newblock Discrete diffusion modeling by estimating the ratios of the data
  distribution.
\newblock In \emph{International Conference on Machine Learning}, 2024.

\bibitem[Sahoo et~al.(2024)Sahoo, Arriola, Schiff, Gokaslan, Marroquin, Chiu,
  Rush, and Kuleshov]{sahoo2024mdlm}
Subham~Sekhar Sahoo, Marianne Arriola, Yair Schiff, Aaron Gokaslan, Edgar
  Marroquin, Justin~T. Chiu, Alexander Rush, and Volodymyr Kuleshov.
\newblock Simple and effective masked diffusion language models.
\newblock In \emph{Advances in Neural Information Processing Systems},
  volume~37, pages 130136--130184, 2024.

\bibitem[Shi et~al.(2024)Shi, Han, Wang, Doucet, and Titsias]{shi2024md4}
Jiaxin Shi, Kehang Han, Zhe Wang, Arnaud Doucet, and Michalis~K. Titsias.
\newblock Simplified and generalized masked diffusion for discrete data.
\newblock In \emph{Advances in Neural Information Processing Systems}, 2024.

\bibitem[Nie et~al.(2025)Nie, Zhu, You, Zhang, Ou, Hu, Zhou, Lin, Wen, and
  Li]{nie2025llada}
Shen Nie, Fengqi Zhu, Zebin You, Xiaolu Zhang, Jingyang Ou, Jun Hu, Jun Zhou,
  Yankai Lin, Ji-Rong Wen, and Chongxuan Li.
\newblock Large language diffusion models.
\newblock \emph{arXiv preprint arXiv:2502.09992}, 2025.

\bibitem[Ye et~al.(2025{\natexlab{b}})Ye, Xie, Zheng, Gao, Wu, Jiang, Li, and
  Kong]{ye2025dream}
Jiacheng Ye, Zhihui Xie, Lin Zheng, Jiahui Gao, Zirui Wu, Xin Jiang, Zhenguo
  Li, and Lingpeng Kong.
\newblock Dream {7B}: Diffusion large language models.
\newblock \emph{arXiv preprint arXiv:2508.15487}, 2025{\natexlab{b}}.

\bibitem[{Inception Labs} et~al.(2025){Inception Labs}, Khanna, Kharbanda, Li,
  et~al.]{inception2025mercury}
{Inception Labs}, Samar Khanna, Siddhant Kharbanda, Shufan Li, et~al.
\newblock Mercury: Ultra-fast language models based on diffusion.
\newblock \emph{arXiv preprint arXiv:2506.17298}, 2025.

\bibitem[Song et~al.(2025{\natexlab{b}})Song, Zhang, Luo, Gao, Xia,
  et~al.]{song2025seeddiffusion}
Yuxuan Song, Zheng Zhang, Cheng Luo, Pengyang Gao, Fan Xia, et~al.
\newblock Seed diffusion: A large-scale diffusion language model with
  high-speed inference.
\newblock \emph{arXiv preprint arXiv:2508.02193}, 2025{\natexlab{b}}.

\bibitem[Bie et~al.(2025)Bie, Cao, Chen, Du, et~al.]{antgroup2025llada2}
Tiwei Bie, Maosong Cao, Kun Chen, Lun Du, et~al.
\newblock {LLaDA2.0}: Scaling up diffusion language models to {100B}.
\newblock \emph{arXiv preprint arXiv:2512.15745}, 2025.

\bibitem[Arriola et~al.(2025)Arriola, Gokaslan, Chiu, Yang, Qi, Han, Sahoo, and
  Kuleshov]{arriola2025bd3lm}
Marianne Arriola, Aaron Gokaslan, Justin~T. Chiu, Zhihan Yang, Zhixuan Qi,
  Jiaqi Han, Subham~Sekhar Sahoo, and Volodymyr Kuleshov.
\newblock Block diffusion: Interpolating between autoregressive and diffusion
  language models.
\newblock In \emph{International Conference on Learning Representations}, 2025.

\bibitem[Gong et~al.(2025)Gong, Agarwal, Zhang, Ye, Zheng, Li, An,
  et~al.]{gong2025diffullama}
Shansan Gong, Shivam Agarwal, Yizhe Zhang, Jiacheng Ye, Lin Zheng, Mukai Li,
  Chenxin An, et~al.
\newblock Scaling diffusion language models via adaptation from autoregressive
  models.
\newblock In \emph{International Conference on Learning Representations}, 2025.

\bibitem[Wu et~al.(2025)Wu, Zhang, Xue, Diao, et~al.]{wu2025fastdllmv2}
Chengyue Wu, Hao Zhang, Shuchen Xue, Shizhe Diao, et~al.
\newblock Fast-{dLLM} v2: Efficient block-diffusion {LLM}.
\newblock \emph{arXiv preprint arXiv:2509.26328}, 2025.

\bibitem[Liu et~al.(2025)Liu, Cao, Li, Luo, Chen, et~al.]{opengvlab2025sdlm}
Yangzhou Liu, Yue Cao, Hao Li, Gen Luo, Zhe Chen, et~al.
\newblock Sequential diffusion language models.
\newblock \emph{arXiv preprint arXiv:2509.24007}, 2025.

\bibitem[{StepFun-Audio Team}(2026)]{stepfun2026stepaudio25}
{StepFun-Audio Team}.
\newblock {StepAudio 2.5} technical report.
\newblock \emph{arXiv preprint arXiv:2605.23463}, 2026.

\bibitem[Fan et~al.(2026)Fan, Xie, Zou, and Chen]{fan2026lladatts}
Xiaoyu Fan, Huizhi Xie, Wei Zou, and Yunzhang Chen.
\newblock {LLaDA-TTS}: Unifying speech synthesis and zero-shot editing via
  masked diffusion modeling.
\newblock \emph{arXiv preprint arXiv:2603.26364}, 2026.

\bibitem[Lou et~al.(2026)Lou, Wu, Wang, et~al.]{lou2026diffuspeech}
Yuxuan Lou, Ziming Wu, Yaochen Wang, et~al.
\newblock {DiffuSpeech}: Silent thought, spoken answer via unified speech-text
  diffusion.
\newblock \emph{arXiv preprint arXiv:2601.22889}, 2026.

\bibitem[Zhu et~al.(2026)Zhu, Ye, Kang, et~al.]{zhu2026omnivoice}
Han Zhu, Lingxuan Ye, Wei Kang, et~al.
\newblock {OmniVoice}: Towards omnilingual zero-shot text-to-speech with
  diffusion language models.
\newblock \emph{arXiv preprint arXiv:2604.00688}, 2026.

\bibitem[Seo et~al.(2026)Seo, Park, and Nam]{seo2026chatterboxflash}
Deokjin Seo, Gangin Park, and Kihyun Nam.
\newblock Chatterbox-flash: Prior-calibrated block diffusion for streaming
  zero-shot {TTS}.
\newblock \emph{arXiv preprint arXiv:2605.30748}, 2026.

\bibitem[Shao et~al.(2024)Shao, Wang, Zhu, Xu, Song, Bi, Zhang, Zhang, Li, Wu,
  and Guo]{shao2024deepseekmath}
Zhihong Shao, Peiyi Wang, Qihao Zhu, Runxin Xu, Junxiao Song, Xiao Bi, Haowei
  Zhang, Mingchuan Zhang, Y.~K. Li, Y.~Wu, and Daya Guo.
\newblock {DeepSeekMath}: Pushing the limits of mathematical reasoning in open
  language models, 2024.
\newblock URL \url{https://arxiv.org/abs/2402.03300}.
\newblock arXiv preprint arXiv:2402.03300.

\bibitem[Zhou et~al.(2025{\natexlab{b}})Zhou, Zeng, Liu, Li, Yu, Wang, Ye, Sun,
  Gui, Li, Wu, and Liu]{zhou2025voxcpm}
Yixuan Zhou, Guoyang Zeng, Xin Liu, Xiang Li, Renjie Yu, Ziyang Wang, Runchuan
  Ye, Weiyue Sun, Jiancheng Gui, Kehan Li, Zhiyong Wu, and Zhiyuan Liu.
\newblock {VoxCPM}: Tokenizer-free {TTS} for context-aware speech generation
  and true-to-life voice cloning.
\newblock \emph{arXiv preprint arXiv:2509.24650}, 2025{\natexlab{b}}.

\bibitem[Ye et~al.(2024)Ye, Sun, Lei, Lin, Tan, Dai, Kong, Chen, Pan, Liu, Guo,
  and Xue]{ye2024xcodec}
Zhen Ye, Peiwen Sun, Jiahe Lei, Hongzhan Lin, Xu~Tan, Zheqi Dai, Qiuqiang Kong,
  Jianyi Chen, Jiahao Pan, Qifeng Liu, Yike Guo, and Wei Xue.
\newblock Codec does matter: Exploring the semantic shortcoming of codec for
  audio language model.
\newblock \emph{arXiv preprint arXiv:2408.17175}, 2024.

\bibitem[Zhang et~al.(2024)Zhang, Zhang, Li, Zhou, and
  Qiu]{zhang2024speechtokenizer}
Xin Zhang, Dong Zhang, Shimin Li, Yaqian Zhou, and Xipeng Qiu.
\newblock {SpeechTokenizer}: Unified speech tokenizer for speech large language
  models.
\newblock In \emph{International Conference on Learning Representations}, 2024.

\bibitem[Chen et~al.(2022)Chen, Wang, Chen, Wu, Liu, Chen, Li, Kanda, Yoshioka,
  Xiao, Wu, Zhou, Ren, Qian, Qian, Wu, Zeng, Yu, and Wei]{chen2022wavlm}
Sanyuan Chen, Chengyi Wang, Zhengyang Chen, Yu~Wu, Shujie Liu, Zhuo Chen, Jinyu
  Li, Naoyuki Kanda, Takuya Yoshioka, Xiong Xiao, Jian Wu, Long Zhou, Shuo Ren,
  Yanmin Qian, Yao Qian, Jian Wu, Michael Zeng, Xiangzhan Yu, and Furu Wei.
\newblock {WavLM}: Large-scale self-supervised pre-training for full stack
  speech processing.
\newblock \emph{IEEE Journal of Selected Topics in Signal Processing},
  16\penalty0 (6):\penalty0 1505--1518, 2022.

\bibitem[Kong et~al.(2020)Kong, Kim, and Bae]{kong2020hifigan}
Jungil Kong, Jaehyeon Kim, and Jaekyoung Bae.
\newblock {HiFi-GAN}: Generative adversarial networks for efficient and high
  fidelity speech synthesis.
\newblock In \emph{Advances in Neural Information Processing Systems}, 2020.

\bibitem[Lee et~al.(2023)Lee, Ping, Ginsburg, Catanzaro, and
  Yoon]{lee2023bigvgan}
Sang-gil Lee, Wei Ping, Boris Ginsburg, Bryan Catanzaro, and Sungroh Yoon.
\newblock {BigVGAN}: A universal neural vocoder with large-scale training.
\newblock In \emph{International Conference on Learning Representations}, 2023.

\bibitem[{Boson AI}(2025)]{bosonai2025higgsaudio}
{Boson AI}.
\newblock Higgs audio v2: Text-audio foundation model.
\newblock \url{https://github.com/boson-ai/higgs-audio}, 2025.
\newblock Open release with a unified semantic--acoustic audio tokenizer; no
  accompanying arXiv report.

\bibitem[Chang et~al.(2022)Chang, Zhang, Jiang, Liu, and
  Freeman]{chang2022maskgit}
Huiwen Chang, Han Zhang, Lu~Jiang, Ce~Liu, and William~T. Freeman.
\newblock {MaskGIT}: Masked generative image transformer.
\newblock In \emph{IEEE/CVF Conference on Computer Vision and Pattern
  Recognition}, 2022.

\bibitem[Zhao et~al.(2025)Zhao, Gupta, Zheng, and Grover]{zhao2025d1}
Siyan Zhao, Devaansh Gupta, Qinqing Zheng, and Aditya Grover.
\newblock d1: Scaling reasoning in diffusion large language models via
  reinforcement learning.
\newblock In \emph{Advances in Neural Information Processing Systems},
  volume~38, pages 56729--56762, 2025.
\newblock URL
  \url{https://proceedings.neurips.cc/paper_files/paper/2025/file/52190a0362148d179f1cbd9080956872-Paper-Conference.pdf}.

\bibitem[Black et~al.(2024)Black, Janner, Du, Kostrikov, and
  Levine]{black2024ddpo}
Kevin Black, Michael Janner, Yilun Du, Ilya Kostrikov, and Sergey Levine.
\newblock Training diffusion models with reinforcement learning.
\newblock In \emph{International Conference on Learning Representations}, 2024.
\newblock URL
  \url{https://proceedings.iclr.cc/paper_files/paper/2024/file/14f75513f0f1ca01de1e826b52e6b840-Paper-Conference.pdf}.

\bibitem[Raajesh et~al.(2026)Raajesh, Shah, Klivans, and
  Kr{\"a}henb{\"u}hl]{raajesh2026maskaware}
Haran Raajesh, Kulin Shah, Adam Klivans, and Philipp Kr{\"a}henb{\"u}hl.
\newblock Mask-aware policy gradients for diffusion language models.
\newblock \emph{arXiv preprint arXiv:2607.15200}, 2026.
\newblock \doi{10.48550/arXiv.2607.15200}.
\newblock Accepted at COLM 2026.

\bibitem[Ho and Salimans(2022)]{ho2022classifierfree}
Jonathan Ho and Tim Salimans.
\newblock Classifier-free diffusion guidance.
\newblock \emph{arXiv preprint arXiv:2207.12598}, 2022.
\newblock \doi{10.48550/arXiv.2207.12598}.

\bibitem[Schulman et~al.(2017)Schulman, Wolski, Dhariwal, Radford, and
  Klimov]{schulman2017ppo}
John Schulman, Filip Wolski, Prafulla Dhariwal, Alec Radford, and Oleg Klimov.
\newblock Proximal policy optimization algorithms.
\newblock \emph{arXiv preprint arXiv:1707.06347}, 2017.
\newblock \doi{10.48550/arXiv.1707.06347}.

\bibitem[Yin et~al.(2026)Yin, Zhu, Gao, Zheng, Huang, Zhou, Yang, Liu, Chen,
  Guo, Deng, Mo, Wang, Cheng, Wang, and Liu]{yin2026vllmomni}
Peiqi Yin, Jiangyun Zhu, Han Gao, Chenguang Zheng, Yongxiang Huang, Taichang
  Zhou, Ruirui Yang, Weizhi Liu, Weiqing Chen, Canlin Guo, Didan Deng, Zifeng
  Mo, Cong Wang, James Cheng, Roger Wang, and Hongsheng Liu.
\newblock {vLLM-Omni}: Fully disaggregated serving for any-to-any multimodal
  models.
\newblock \emph{arXiv preprint arXiv:2602.02204}, 2026.
\newblock \doi{10.48550/arXiv.2602.02204}.

\bibitem[Zheng et~al.(2025)Zheng, Chen, Niu, Ma, Wang, Yu, and
  Chen]{zheng2025fastf5latency}
Qixi Zheng, Yushen Chen, Zhikang Niu, Ziyang Ma, Xiaofei Wang, Kai Yu, and Xie
  Chen.
\newblock Accelerating flow-matching-based text-to-speech via empirically
  pruned step sampling.
\newblock \emph{arXiv preprint arXiv:2505.19931}, 2025.
\newblock \doi{10.48550/arXiv.2505.19931}.

\bibitem[Zhu et~al.(2025)Zhu, Kang, Yao, Guo, Kuang, Li, Zhuang, Lin, and
  Povey]{zhu2025zipvoicelatency}
Han Zhu, Wei Kang, Zengwei Yao, Liyong Guo, Fangjun Kuang, Zhaoqing Li, Weiji
  Zhuang, Long Lin, and Daniel Povey.
\newblock Zipvoice: Fast and high-quality zero-shot text-to-speech with flow
  matching.
\newblock \emph{arXiv preprint arXiv:2506.13053}, 2025.
\newblock \doi{10.48550/arXiv.2506.13053}.

\bibitem[{OpenBMB}(2026)]{openbmb2026voxcpm2latency}
{OpenBMB}.
\newblock {VoxCPM2}: Production deployment and inference performance.
\newblock \url{https://github.com/OpenBMB/VoxCPM}, 2026.
\newblock Accessed: 2026-08-10.

\bibitem[{SparkAudio}(2025)]{sparkaudio2025sparkttslatency}
{SparkAudio}.
\newblock {Spark-TTS}: Nvidia triton inference serving.
\newblock \url{https://github.com/SparkAudio/Spark-TTS}, 2025.
\newblock Accessed: 2026-08-10.

\bibitem[{SGLang Team}(2026)]{sglangomni2026}
{SGLang Team}.
\newblock {SGLang-Omni}: High-performance multi-stage pipeline framework for
  omni models.
\newblock \url{https://github.com/sgl-project/sglang-omni}, 2026.
\newblock Accessed: 2026-08-10.

\bibitem[{Cartesia}(2026)]{cartesia2026sonic35latency}
{Cartesia}.
\newblock Sonic 3.5 self-hosted hardware selection and latency.
\newblock \url{https://docs.cartesia.ai/self-hosted/hardware-selection}, 2026.
\newblock Accessed: 2026-08-10.

\bibitem[{ElevenLabs}(2026{\natexlab{a}})]{elevenlabs2026flashv25latency}
{ElevenLabs}.
\newblock Models: Eleven flash v2.5.
\newblock \url{https://elevenlabs.io/docs/overview/models}, 2026{\natexlab{a}}.
\newblock Accessed: 2026-08-10.

\bibitem[{Hume AI}(2026)]{hume2026octave2latency}
{Hume AI}.
\newblock Text-to-speech: Octave 2.
\newblock \url{https://dev.hume.ai/docs/text-to-speech-tts/overview}, 2026.
\newblock Accessed: 2026-08-10.

\bibitem[{PlayHT}(2026)]{playht2026play30minilatency}
{PlayHT}.
\newblock Text-to-speech models: Play 3.0 mini.
\newblock \url{https://docs.play.ht/reference/models}, 2026.
\newblock Accessed: 2026-08-10.

\bibitem[{Deepgram}(2025)]{deepgram2025aura2latency}
{Deepgram}.
\newblock Aura-2 text-to-speech performance.
\newblock \url{https://developers.deepgram.com/changelog/2025/5/14}, 2025.
\newblock Accessed: 2026-08-10.

\bibitem[{Microsoft}(2025)]{microsoft2025vibevoicerealtime}
{Microsoft}.
\newblock Vibevoice-realtime: Real-time streaming text-to-speech.
\newblock \url{https://github.com/microsoft/VibeVoice}, 2025.
\newblock Accessed: 2026-08-10.

\bibitem[Zhou et~al.(2026)Zhou, Zeng, Liu, Li, Yu, Gui, Wu, Wang, Shen, Ye,
  Zhang, Zhou, Bai, Sun, Deng, Shi, Wu, and Liu]{zhou2026voxcpm2}
Yixuan Zhou, Guoyang Zeng, Xin Liu, Xiang Li, Renjie Yu, Jiancheng Gui, Jiaheng
  Wu, Ziyang Wang, Xudong Shen, Runchuan Ye, Zhisheng Zhang, Jiuyang Zhou,
  Bingsong Bai, Weiyue Sun, Mengyuan Deng, Qundong Shi, Zhiyong Wu, and Zhiyuan
  Liu.
\newblock {VoxCPM2} technical report.
\newblock \emph{arXiv preprint arXiv:2606.06928}, 2026.

\bibitem[{MiniMax}(2026)]{minimax2026speech28}
{MiniMax}.
\newblock {MiniMax Speech-2.8}.
\newblock \url{https://www.minimax-speech.com/}, 2026.
\newblock Accessed: 2026-07-30.

\bibitem[{ElevenLabs}(2026{\natexlab{b}})]{elevenlabs2026elevenv3}
{ElevenLabs}.
\newblock {Eleven v3}.
\newblock \url{https://elevenlabs.io/v3}, 2026{\natexlab{b}}.
\newblock Accessed: 2026-07-30.

\bibitem[{Fish Audio}(2026)]{fishaudio2026s21}
{Fish Audio}.
\newblock Introducing {S2.1 Pro}: Our most expressive {TTS} model yet.
\newblock \url{https://fish.audio/blog/s2-1-pro-free-api/}, 2026.
\newblock Accessed: 2026-07-30.

\bibitem[Ni et~al.(2026)Ni, Liao, Chen, Wang, and Wu]{ni2026nvbench}
Qinke Ni, Huan Liao, Dekun Chen, Yuxiang Wang, and Zhizheng Wu.
\newblock {NV-Bench}: Benchmark of nonverbal vocalization synthesis for
  expressive text-to-speech generation.
\newblock \emph{arXiv preprint arXiv:2603.15352}, 2026.
\newblock \doi{10.48550/arXiv.2603.15352}.

\bibitem[Xue et~al.(2026)Xue, Bian, Pan, Wu, Ren, Kang, Hu, Ma, Wang, Qian,
  et~al.]{xue2026nvvsuperbench}
Liumeng Xue, Weizhen Bian, Jiahao Pan, Wenxuan Wu, Yilin Ren, Boyi Kang,
  Jingbin Hu, Ziyang Ma, Shuai Wang, Xinyuan Qian, et~al.
\newblock {NVV-SuperBench}: Beyond words, beyond quality---benchmarking
  nonverbal vocalizations in speech generation.
\newblock \emph{arXiv preprint arXiv:2604.16211}, 2026.
\newblock \doi{10.48550/arXiv.2604.16211}.

\bibitem[{Google DeepMind}(2026)]{google2026gemini31pro}
{Google DeepMind}.
\newblock {Gemini 3.1 Pro} model card.
\newblock \url{https://deepmind.google/models/model-cards/gemini-3-1-pro/},
  2026.
\newblock Accessed: 2026-07-30.

\bibitem[Zhou et~al.(2022)Zhou, Sisman, Liu, and Li]{zhou2022emotional}
Kun Zhou, Berrak Sisman, Rui Liu, and Haizhou Li.
\newblock Emotional voice conversion: Theory, databases and {ESD}.
\newblock \emph{Speech Communication}, 137:\penalty0 1--18, 2022.
\newblock \doi{10.1016/j.specom.2021.11.006}.

\bibitem[Ma et~al.(2024)Ma, Zheng, Ye, Li, Gao, Zhang, and
  Chen]{ma2024emotion2vec}
Ziyang Ma, Zhisheng Zheng, Jiaxin Ye, Jinchao Li, Zhifu Gao, Shiliang Zhang,
  and Xie Chen.
\newblock emotion2vec: Self-supervised pre-training for speech emotion
  representation.
\newblock In \emph{Findings of the Association for Computational Linguistics:
  ACL 2024}, pages 15747--15760, 2024.
\newblock \doi{10.18653/v1/2024.findings-acl.931}.

\bibitem[Manku et~al.(2025)Manku, Tang, Shi, Li, and
  Smola]{manku2025emergentttseval}
Ruskin~Raj Manku, Yuzhi Tang, Xingjian Shi, Mu~Li, and Alexander~J. Smola.
\newblock Emergenttts-eval: Evaluating tts models on complex prosodic,
  expressiveness, and linguistic challenges using model-as-a-judge.
\newblock In \emph{Advances in Neural Information Processing Systems},
  volume~38, 2025.

\bibitem[Chen et~al.(2026)Chen, Hu, Xue, Zhan, Li, Ma, Xie, Guo, Ma, Jiang, Wu,
  Xie, Xie, Zhang, and Xie]{chen2026mintbench}
Huakang Chen, Jingbin Hu, Liumeng Xue, Qirui Zhan, Wenhao Li, Guobin Ma, Hanke
  Xie, Dake Guo, Linhan Ma, Yuepeng Jiang, Bengu Wu, Pengyuan Xie, Chuan Xie,
  Qiang Zhang, and Lei Xie.
\newblock Mint-bench: A comprehensive multilingual benchmark for
  instruction-following text-to-speech.
\newblock \emph{arXiv preprint arXiv:2604.17958}, 2026.

\end{thebibliography}

\end{document}